\documentclass[preprint,authoryear,12pt]{elsarticle}

\pdfoutput=1

\usepackage{graphicx}
\usepackage{epstopdf}
\usepackage{subfigure}
\usepackage{float}
\usepackage{esvect}
\usepackage{amsmath}
\usepackage{amssymb}
\usepackage{commath}
\usepackage[plainpages=false,pdfpagelabels,hypertexnames=false,bookmarks=false]{hyperref}
\hypersetup{pdfborder={0 0 0}}
\usepackage[authoryear,round]{natbib}

\makeatletter
\def\ps@pprintTitle{%
 \let\@oddhead\@empty
 \let\@evenhead\@empty
 \def\@oddfoot{}%
 \let\@evenfoot\@oddfoot}
\makeatother

\journal{Journal of Marine Systems}

\begin{document}

\renewcommand\figref{Fig.~\ref}

\begin{frontmatter}

\title{The influence of winter convection on primary production: a parameterisation using a hydrostatic three-dimensional biogeochemical model}

\author[addLabel:WR]{Fabian Gro{\ss}e\corref{CorrAuthor}} \ead{fabian.grosze@informatik.uni-hamburg.de}
\author[addLabel:DTU]{Christian Lindemann} \ead{chrli@aqua.dtu.dk}
\author[addLabel:IfM]{Johannes P{\"a}tsch} \ead{johannes.paetsch@zmaw.de}
\author[addLabel:IfM]{Jan O. Backhaus} \ead{jan.backhaus@zmaw.de}

\address[addLabel:WR]{Research Group Scientific Computing, Department of Informatics, University of Hamburg, Bundesstra{\ss}e 45a, 20146 Hamburg, Germany}
\address[addLabel:DTU]{National Institute of Aquatic Resources, Technical University of Denmark, J{\ae}gersborg Alle 1, 2920 Charlottenlund, Denmark}
\address[addLabel:IfM]{Institute of Oceanography, University of Hamburg, Bundesstra{\ss}e 53, 20146 Hamburg, Germany}

\cortext[CorrAuthor]{Corresponding author: Fabian Gro{\ss}e, Phone: 0049 40 460 094 115} 

\begin{abstract}
\indent In the recent past observational and modelling studies have shown that the vertical displacement of water parcels, and therefore, phytoplankton particles in regions of deep-reaching convection plays a key role in late winter/early spring primary production. The underlying mechanism describes how convection cells capture living phytoplankton cells and recurrently expose them to sunlight.\\
\indent This study presents a parameterisation called `phytoconvection' which focuses on the influence of convection on primary production. This parameterisation was implemented into a three-dimensional physical-biogeochemical model and applied to the Northwestern European Continental Shelf and areas of the adjacent Northeast Atlantic. The simulation was compared to a `conventional' parameterisation with respect to its influence on phytoplankton concentrations during the annual cycle and its effect on the carbon cycle.\\
\indent The simulation using the new parameterisation showed good agreement with observation data recorded during winter, whereas the reference simulation did not capture the observed phytoplankton concentrations. The new parameterisation had a strong influence on the carbon export through the sinking of particulate organic carbon. The carbon export during late winter/early spring significantly exceeded the export of the reference run.\\
\indent Furthermore, a non-hydrostatic convection model was used to evaluate the major assumption of the presented parameterisation which implies the matching of the mixed layer depth with the convective mixing depth. The applied mixed layer depth criterion principally overestimates the actual convective mixing depth. However, the results showed that this assumption is reasonable during late winter, while indicating a mismatch during spring.
\end{abstract}

\begin{keyword}
Convection \sep Primary production \sep Parameterisation \sep Biogeochemical modelling \sep Northeast Atlantic \sep Northwestern European Continental Shelf
\end{keyword}

\fntext[abbreviations]{Abbreviations: 3D -- three-dimensional, ASCF -- air-sea carbon flux, CEP -- carbon export production, LLF -- light limitation function, MLD -- mixed layer depth, NECS -- Northwestern European Continental Shelf, NPP -- net primary production}

\end{frontmatter}

\section{Introduction} \label{sec:Intro}
\indent Photosynthesis in the ocean has been estimated to contribute almost 50\% to the total global net primary production \citep{field1998} and total carbon uptake \citep{sabine2004}. This large stake of marine primary production shows the great importance of understanding the physical, chemical and biological processes which influence marine primary production. Marine primary production is generally controlled by the availability of light and nutrients. In the ocean, light decreases exponentially with depth, limiting photosynthesis to the upper part of the ocean, the euphotic zone.\\
\indent The `critical depth model' \citep{sverdrup1953} describes the relation between the depth of the surface mixed layer and the capability of light-dependent phytoplankton net growth. It defines the compensation depth as the depth where the gain, or growth, and the loss in phytoplankton balance each other. Hence, there exists a critical depth where the vertically integrated growth is equal to the vertically integrated loss. \citet{sverdrup1953} concluded that net phytoplankton growth is only possible if the mixed layer depth (MLD) is less than the critical depth, thus, allowing for positive net growth.\\
\indent During winter the MLD in the North Atlantic reaches depths of several hundred metres \citep[e.g.][]{mccartney1982} and therefore would not allow net phytoplankton growth according to the `critical depth model', that is unless the loss terms are sufficiently low, and thus, result in a sufficiently deep critical depth (critical depth $\geq$ MLD). In this context, \citet{behrenfeld2010} proposed the so-called `dilution-recoupling hypothesis', which based upon earlier work by \citet{evans1985}, arguing that the dilution of phytoplankton by mixed layer deepening reduces the grazing pressure on phytoplankton, and thus, lowers this loss term during winter. This hypothesis has been updated to the more general `disturbance-recoupling hypothesis' \citep{behrenfeld2013, behrenfeld2014} stating that during the annual cycle the predator-prey interaction is disrupted by environmental factors as, for example, convection, and subsequently recovers. This disruption then allows for net primary production due to low grazing pressure while concentrations still stay low due to convective mixing.\\
%
\begin{figure}[H]
\centering
\includegraphics[width=0.4\textwidth]{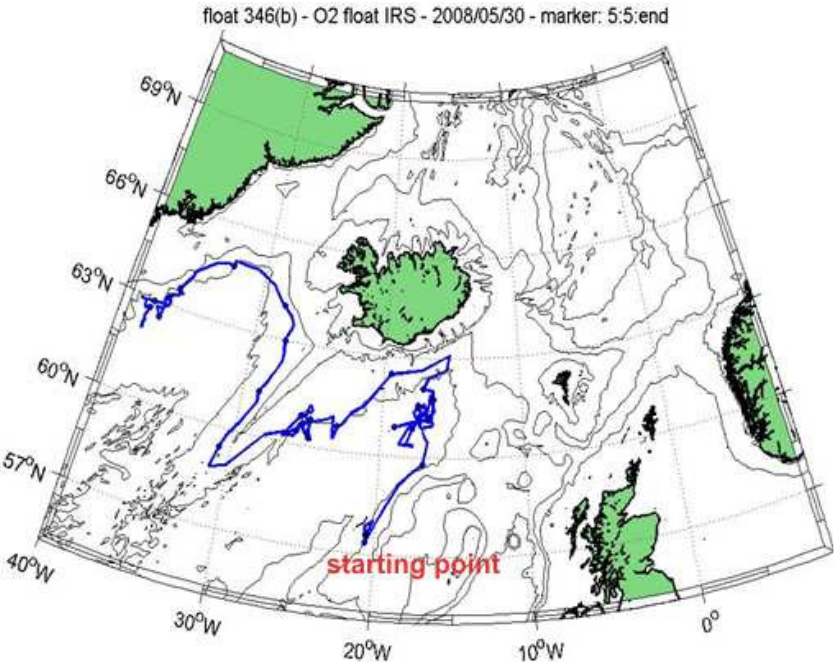}
\caption{Trajectory of the ARGO float which recorded the time series from September 2005 to May 2008 shown in \figref{fig:ARGO_Hov}. Source: D. Quadfasel, Institute of Oceanography, University of Hamburg (unpublished data).}
\label{fig:ARGO_Tra}
\end{figure}
%
\indent Ocean convection -- the buoyancy-driven sinking of surface waters due to surface cooling or an increase in surface salinity -- is one of the key processes affecting the winter mixed layer deepening. Due to mass conservation the sinking of water parcels leads to a balancing upward motion, and thus, convection can be described as an orbital motion. Especially in the North Atlantic Ocean extensive convectively driven mixed layer deepening can be observed during winter. \citet{mccartney1982} determined an average `late winter' (January - April) MLD of more than 400\,m over large parts of the subpolar North Atlantic using temperature and salinity profiles recorded during the 1950s and 1960s. \citet{holliday2000} reported on convection in the Rockall Trough removing the seasonal thermocline and reaching down to depths of around 700\,m.\\ 
\indent In the meantime studies showed that convection plays a key role for primary production during winter \citep[e.g.][]{backhaus1999,wehde2001,ward2007}. \citet{backhaus1999} hypothesised a direct link between ocean convection and winter primary production called `phytoconvection'. They suggested that the upward and downward motion within a convection cell causes phytoplankton particles to regularly re-enter the euphotic zone allowing them to grow, and thus, balancing their losses due to respiration, mortality, grazing and sinking. This relation was supported by modelling and observational studies \citep{wehde2000,wehde2001,backhaus2003}. \citet{ward2007} emphasise the role of convective upward motion which counteracts the sinking of phytoplankton, and thus, increases net growth.\\
%
\begin{figure}[H]
\centering
\includegraphics[width=0.6\textwidth]{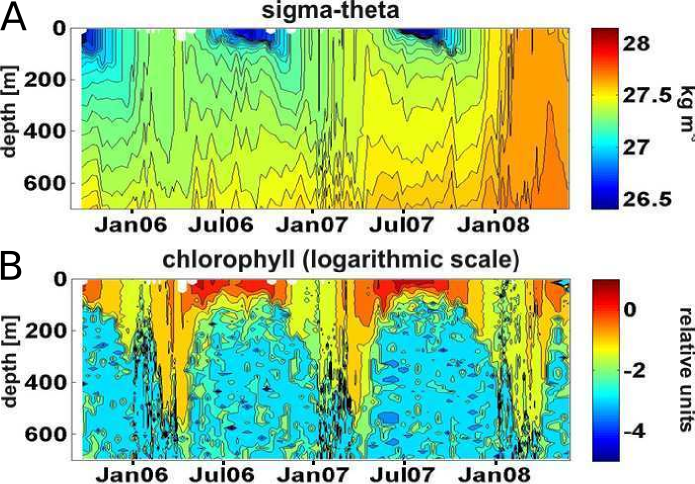}
\caption{Hovm{\"o}ller diagrams of (A) derived potential density anomaly $\sigma_\theta$ in kg\,m$^{-3}$ and (B) chlorophyll derived from fluorescence data in non-calibrated relative units. Source: Quadfasel (unpublished data).}
\label{fig:ARGO_Hov}
\end{figure}
\indent \figref{fig:ARGO_Hov} shows the time series of potential density anomaly $\sigma_\theta$ and chlorophyll derived from fluorescence data. The time series was recorded by an ARGO float released in the eastern Iceland Basin in September 2005 and then drifted along the topography of the Reykjanes Ridge into the Irminger Basin until May 2008 (see \figref{fig:ARGO_Tra}). The time series of $\sigma_\theta$ (A) shows a distinct surface mixed layer in autumn 2005 and from spring to autumn in 2006 and 2007. Due to the different hydrodynamics in the Irminger Basin compared to the Iceland Basin \citep[e.g.][]{krauss1995} no stratification is visible in spring 2008. During late winter 2006 and 2007 the greatest MLD reached up to 600 m. The chlorophyll time series (B) is presented in relative units due to a lack of calibration data. However, in the chlorophyll the same patterns as in $\sigma_\theta$ are shown underlining the importance of winter convective mixing for the phytoplankton community.\\
\indent In large-scale ocean models, an adequate representation of the influence of convection on primary production is often missing. These models usually work on large horizontal and smaller vertical length scales, and therefore normally neglect the vertical acceleration. For many large- and mesoscale processes this does not cause any restrictions and the reduction of the vertical resolution saves computation time. However, when dealing with motions characterised by spatial aspect ratios of order $1$, like convection, the hydrostatic approximation becomes inaccurate \citep{marshall1998}. Furthermore, the coarse vertical resolution of these models cannot resolve the upward and downward motion of water parcels with high vertical velocities as it is known for ocean convection.\\
\indent As a consequence of these two restrictions, models which do not a priori consider all variables to be homogeneously distributed within the mixed layer, may not represent the vertical exchange of water mass properties and phytoplankton in deep MLDs properly. This necessitates a parameterisation of primary production which is able to reproduce observed winter phytoplankton using these models \citep{holt2014}.\\
\indent Previous parameterisations of the influence of convection on primary production in Eulerian models lead to a reduction of primary production in deep MLDs \citep[e.g.][]{levy1998a,levy1998b}. Furthermore, these studies focused on the onset of the bloom in spring after winter convection \citep{levy1998b} or did not deal with mixed layers deeper than 200\,m \citep{levy1998a} while the study at hand focusses on primary production during winter when convection is present reaching depths of up to 500\,m. \citet{backhaus2003} suggested an approach for the parameterisation of phytoconvection taking into account the spatial aspect ratio of convective orbits. \citet{janout2003} adopted the idea and implemented it in a one-dimensional hydrostatic physical-biological model. The idea behind his approach was to compensate the lack of convective vertical displacement of phytoplankton by allowing primary production throughout the whole convective mixed layer. That was done by distributing the daily averaged surface solar radiation over the whole mixed layer during winter. To account for summer situations in which the `conventional' approach following \citet{sverdrup1953} is more applicable, he switched between the conventional parameterisation and the phytoconvective approach depending on the MLD. This switching was performed abruptly by applying only the conventional parameterisation during periods with a MLD less than 75\,m and applying the phytoconvective type for deeper mixed layers.\\
\indent The present study incorporates the parameterisation of \citet{janout2003} and aims for the further development of this approach with two major objectives. First, to develop a smoother transition scheme between the conventional and the phytoconvective parameterisation of primary production. Second, to improve the applicability in a large-scale, three-dimensional, physical-biogeochemical model applied to an area including regions of deep winter convection as well as shallow shelf seas.
\section{Material and methods} \label{sec:MatsAndMeth}
\subsection{The three-dimensional physical-biogeochemical model} \label{subsec:PhysBiogeochemModel}
\indent For the three-dimensional (3D) simulations the ECOHAM4 model system \citep[ECOlogical model, HAMburg, version 4;][]{paetsch2008} was used. The model consists of two main modules: the hydrodynamic model HAMSOM \citep[HAMburg Shelf Ocean Model;][]{backhaus1985} and the biogeochemical model ECOHAM. HAMSOM simulates the temperature and salinity distributions, the 3D advective flow field and the turbulent mixing. It is a baroclinic primitive equation model with a free surface and uses the hydrostatic and Boussinesq approximation. HAMSOM is defined on an Arakawa C-grid \citep{arakawa1977} resolving the vertical in z-coordinates. The horizontal advective flow field is calculated using the component upstream scheme. A detailed description of HAMSOM is given by \citet{backhaus1987} and \citet{pohlmann1991,pohlmann1996}. The vertical turbulent mixing is parametrised by the exchange coefficient $A_V$ in depth $z$ assuming stationarity and neglecting advection and diffusion of turbulent kinetic energy \citep{mellor1974}:
\begin{equation}
 A_V(z) = (C_{ml}\cdot H_{ml})^2\cdot\sqrt{\left(\frac{\partial u}{\partial z}\right)^2+\left(\frac{\partial v}{\partial z}\right)^2-\frac{N^2}{S_m}}. \label{equ:VertExchange}
\end{equation}
Here, $u$ and $v$ represent the zonal and meridional velocity components, respectively. $N$ is the Brunt-V{\"a}is{\"a}l{\"a}-frequency and $S_m$ is the Schmidt number. $H_{ml}$, representing the MLD, and $C_{ml}$ are the only terms which must be prescribed. $C_{ml} \approx 0.05 $ is determined after \citet{kochergin1987}. If unstable conditions occur ($N^2<0$), $A_V$ is set to the maximum vertical exchange coefficient $A_{Vmax} = 800\cdot 10^{-4}$\,m$^2$\,s$^{-1}$ to represent the strong vertical mixing due to convective processes which cannot be resolved by HAMSOM.\\
\indent The biogeochemical model ECOHAM describes the cycles of carbon (C), nitrogen (N), phosphorus (P), silicon (Si) and oxygen (O$_2$). It applies a partly variable C:N stoichiometry depending on the state variable and a simple description of benthic processes \citep{paetsch2008}. The model includes the following state variables: four nutrients (nitrate, ammonium, phosphate, silicate), two phytoplankton groups (diatoms and flagellates), two zooplankton groups (micro- and mesozooplankton), bacteria, two fractions of detritus (fast and slowly sinking), labile dissolved organic matter, semi-labile organic carbon, oxygen, calcite, dissolved inorganic carbon and total alkalinity. For a detailed description and a full list of the fluxes between the different state variables and the model parameters, see \citet{lorkowski2012}.
\subsubsection{The parameterisation of `phytoconvection'} \label{subsubsec:Parameterisation}
\indent In ECOHAM the phytoplankton production $P^B$ is controlled by an extended form of `Liebig's law' \citep{liebig1840} considering light and nutrient limitation and accounting for the temperature dependence:
\begin{equation}
 P^B = \gamma \cdot f_T \cdot \min(L,N). \label{equ:liebigMOD}
\end{equation}
The production rate $P^B$ depends on the maximum growth rate $\gamma$, the light $L$ and the nutrients $N$. $f_T$ is a constant factor parameterising the temperature dependence of primary production. For our area of interest and the late winter situation it has been shown that light is the limiting factor during winter as the whole euphotic zone is enriched with nutrients due the upward mixing of nutrient-rich deep water driven by strong surface cooling, and thus, convection. Hence, in the following we will focus on the light dependence of phytoplankton production. The empirical relationship between irradiance and primary production found by \citet{steele1962} builds the basis for the conventional parameterisation of the light-dependent production $P^B$ at depth $z$:
\begin{equation}
 P^B_{Steele}(z)=P^B_{max}\frac{I_{par}(z)}{I_{opt}}\cdot exp\left(1-\frac{I_{par}(z)}{I_{opt}}\right), \label{equ:PB_Steele}
\end{equation}
in which $P^B_{max}$ depicts the maximum biomass-specific production rate ($P^B_{max} = 1.1$\,d$^{-1}$ for diatoms, $P^B_{max} = 0.9$\,d$^{-1}$ for flagellates) and $I_{par}$ is the depth-dependent photosynthetically active radiation (or PAR) and $I_{opt}$ is the dynamically calculated optimal light intensity. 
$I_{par}$ in depth $z$ is calculated as:
\begin{equation}
 I_{par}(z) = k_{par} \cdot I_{sw} \cdot exp(-\epsilon(z)). \label{equ:Ipar}
\end{equation}
Here, $k_{par} = 0.43$ \citep{paetsch2008} represents the photosynthetically active fraction of the incoming short-wave radiation at the sea surface $I_{sw}$. The depth-dependent attenuation coefficient $\epsilon(z)$ is:
\begin{equation}
 \epsilon(z) = (k_w + k_p \cdot C_p + k_s \cdot C_s) \cdot z, \label{equ:AttCoeff}
\end{equation}
and includes the effect of light attenuation by water, planktonic self-shading and turbidity due to suspended particulate matter. $k_w$ is the locally varying attenuation coefficient of water after \citet{jerlov1976}, $k_p = 0.03$ and $C_p$ are the attenuation coefficient in m$^2$\,mmol\,C$^{-1}$ and the concentration in mmol\,C\,m$^{-3}$ of phytoplankton, respectively. $k_s = 0.06$ and $C_s$ are the attenuation coefficient m$^2$\,g$^{-1}$ and the concentration in g\,m$^{-3}$ of suspended particulate matter, respectively. ECOHAM uses a variable $I_{opt}$ and its adaptation to the actual light conditions over time is described in \citet{paetsch2008}. The range of the optimum irradiance $I_{opt}$ is limited to:
\begin{equation}
40 \leq I_{opt} \leq 70 \quad \mbox{[W\,m$^{-2}$]}. \label{equ:IOptRange}
\end{equation}
\indent Starting from equation \eqref{equ:PB_Steele} the new parameterisation of light-dependent primary production under convective conditions is developed. Convection described as the upward and downward motion of water masses is here considered as an orbital system within the mixed layer. \citet{turner1979} and \citet{kaempf1998} studied the spatial and temporal scales of convective cells using observations and numerical simulations, respectively. The spatial aspect ratio (horizontal vs. vertical scale) has been reported to be on average about 2.5:1 \citep{wehde2000} and ranging between 2:1 and 3:1 \citep{kaempf1998}.\\
\indent The convection cell is considered having a rectangular geometry with the depth of the convective mixed layer $H_{cml}$ defining the vertical dimension \citep{backhaus2003}. This represents a simplified approach to account for the spatial aspect ratio of convective orbits, since convective motion is highly turbulent and far from linear movement along a rectangular track.\\
%
\begin{figure}[H]
\centering
\includegraphics[width=0.6\textwidth]{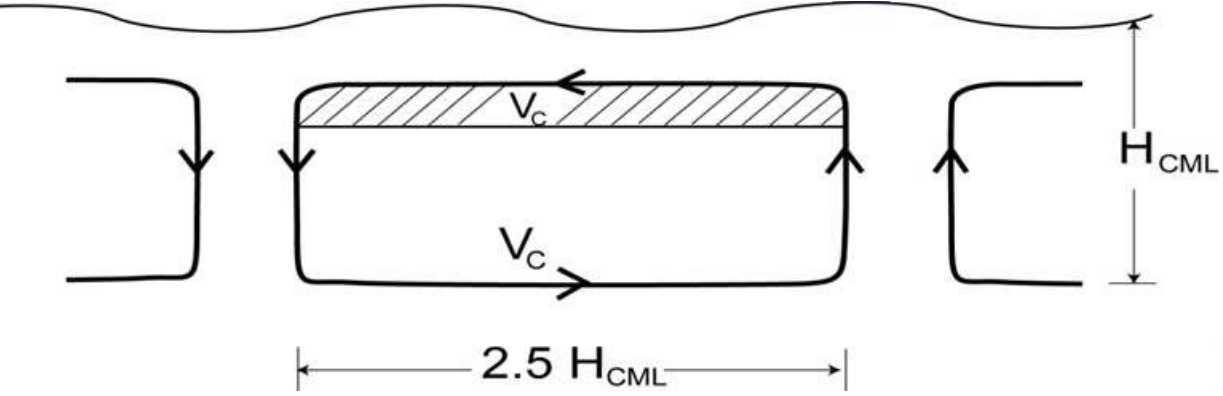}
\caption{Scheme of the convection cell as assumed in the parameterisation of phytoconvection. $H_{cml}$ indicates the convective mixed layer depth, $v_c$ is the velocity of the flow within the convective orbit. The shaded part at the sea surface marks the euphotic zone. (Source: \citet{janout2003})}
\label{fig:conv_cell}
\end{figure}
\indent The conceptual view of such a convection cell is shown in \figref{fig:conv_cell}. It is assumed that during winter the MLD is identical to $H_{cml}$. Hence, the MLD determines the vertical extent of the convection cell. Using the spatial aspect ratio of 2.5:1 and the time of a complete orbit $t_{orb}$ can be calculated as \citep{backhaus2003,janout2003}:
\begin{equation}
 t_{orb}=\frac{H_{cml}}{v_c}+\frac{H_{cml}}{v_c}+2\cdot\frac{2.5 \cdot H_{cml}}{v_c}. \label{equ:Torb}
\end{equation}
Here, $v_c$ is the velocity of the convective motion which is assumed to be constant with $v_c=5$\,cm\,s$^{-1}$.\\
\indent Following \citet{janout2003}, who adapted \citet{backhaus2003} the time within the euphotic zone is defined as:
\begin{equation}
 t_{exp}=\frac{2.5 \cdot H_{cml}}{v_c}+\frac{H_{euph}}{v_c}+\frac{H_{euph}}{v_c} \label{equ:Texp}
\end{equation}
with the euphotic depth $H_{euph}$. $H_{euph}$ is set to the depth where the available light is still 1\% of the surface radiation. $H_{cml}$ is defined as the last depth $z$ in which the temperature difference criterion \eqref{equ:CML_crit} is satisfied.
\begin{equation}
 SST - T(z) \leq 0.4\,\mbox{K}. \label{equ:CML_crit}
\end{equation}
The value of $\Delta T = 0.4$\,K is within the range of literature values varying between $\Delta T = 0.1$\,K and $\Delta T = 1$\,K compared to the sea surface temperature (SST) (see e.g. Table 1 in \citet{kara2000}).\\
\indent Using equations \eqref{equ:Torb}, \eqref{equ:Texp} and the MLD criterion \eqref{equ:CML_crit} the ratio $t_{exp}/t_{orb}$ can be calculated. This ratio implies that each phytoplankton particle within a convective cell has the same probability of residence within the euphotic zone which is the fundamental idea behind the parameterisation presented here. It is assumed that phytoplankton production still follows equation \eqref{equ:PB_Steele} and is conducted in the euphotic zone with the average production rate under the actual light conditions. Combining this assumption with the above ratio $t_{exp}/t_{orb}$, we estimate the `phytoconvective production' $P^B_{pc}$:
\begin{equation}
 P^B_{pc} = \frac{t_{exp}}{t_{orb}} \cdot \frac{1}{H_{euph}} \int_{-H_{euph}}^0 P^B_{Steele}(z)\,dz, \label{equ:PB_pc}
\end{equation}
which is constant throughout the whole mixed layer. It follows that under convective conditions each plankton particle has the same probability to conduct primary production with the average light within the euphotic zone independent of its actual position within the mixed layer.\\
\indent The only term changing in equation \eqref{equ:PB_pc} is the ratio $t_{exp}/t_{orb}$, since equation \eqref{equ:PB_pc} is only applied if $H_{cml} > H_{euph}$ and $P^B_{Steele}$ and $H_{euph}$ are assumed to be identical for different $H_{cml}$. Thus, even though $t_{exp}$ increases with increasing $H_{cml}$, $t_{exp}/t_{orb}$ decreases because of the faster increase of $t_{orb}$. In other words, the frequency of convective orbits is inversely correlated with $H_{cml}$ and the same holds for the exposure to light \citep{backhaus2003}, causing $P^B_{pc} $ to be reduced with increasing $H_{cml}$.\\
\indent Convection of several hundreds of metres depth only occurs during winter, making it necessary to distinguish between convective and non-convective periods. The transition between convective and non-convective regimes is controlled by $H_{cml}$ which is used to build the weighting function \eqref{equ:f_p}. This weighting function controls the influence of the conventional parameterisation $P^B_{Steele}$ (equation \eqref{equ:PB_Steele}) and the phytoconvective parameterisation $P^B_{pc}$ (equation \eqref{equ:PB_pc}). Furthermore, there is a discontinuity between $P^B_{Steele}$ and $P^B_{pc}$ in the case of $H_{cml} = H_{euph}$ due to the additional distance $2.5\cdot H_{cml}$ travelled at the surface (see \figref{fig:conv_cell}). This discontinuity is eluded by the use of the weighting function.
\begin{equation}
f_{p} = \min\left(1,\frac{\max(0, H_{cml}-H_{euph})}{H_{ref}-H_{euph}}\right). \label{equ:f_p}
\end{equation}
This weighting function $f_p$ is set to $0$, if $H_{cml}$ is less than or equal to the euphotic depth $H_{euph}$. Accordingly, $f_p$ is 1, if $H_{cml}$ is equal or greater than the reference depth $H_{ref} = 100$\,m. Thus, $H_{ref}$ defines the limit between purely convective conditions ($H_{cml} \ge H_{ref}$) and transitional conditions ($H_{euph} < H_{cml} < H_{ref}$). The value for $H_{ref}$ was chosen after a test run over one year in which a maximum euphotic depth $H_{euph}$ of about 45\,m occurred. $H_{ref}$ must be set to a significantly higher value than the euphotic depth $H_{euph}$ to allow a steady transition between the conventional parameterisation \eqref{equ:PB_Steele} and the parameterisation of phytoconvection \eqref{equ:PB_pc}.\\
\indent The weighting factor $f_p$ is then applied for the calculation of the actual light-dependent phytoplankton production in depth $z$:
\begin{equation}
P^B_{total}(z) = (1-f_{p}) \cdot P^B_{Steele}(z) + f_{p} \cdot P^B_{pc}. \label{equ:PB_total}
\end{equation}
During summer, when the mixed layer is shallower than the euphotic zone, the phytoconvective parameterisation \eqref{equ:PB_pc} is not taken into account. Conversely, during winter, when the MLD reaches several hundreds of metres the conventional parameterisation \eqref{equ:PB_Steele} is not affecting the phytoplankton productivity. The light-dependent production rate $P^B_{total}$ according to equation \eqref{equ:PB_total} is consequently used for the calculation of primary production within the model.
\subsubsection{Model setup and data} \label{subsubsec:ModelSetup}
\begin{figure}[H]
\centering
\includegraphics[width=0.55\textwidth]{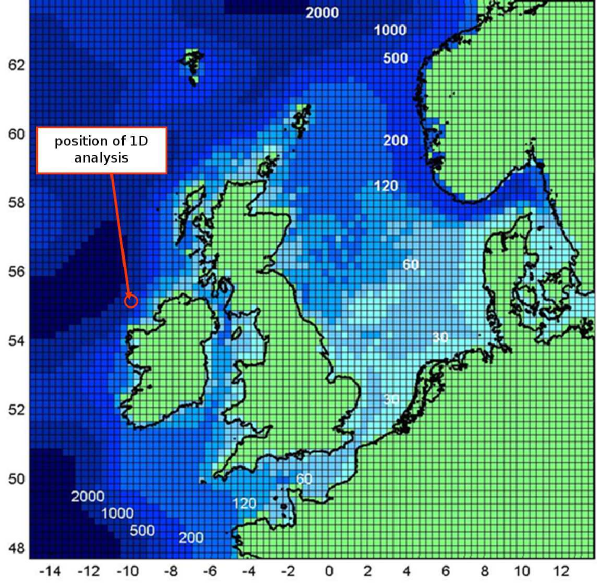}
\caption{Horizontal grid and bottom topography of the NECS area as used by ECOHAM4. The x- and y-axes values mark longitude and latitude, respectively. The red circle marks the station of the one-dimensional analysis (55$^\circ$\,3'\,24''\,N, 10$^\circ$\,14'\,W).}
\label{fig:ModelArea}
\end{figure}

\indent The model was set up on the region of the Northwestern European Continental Shelf (NECS) and parts of the adjacent Northeast Atlantic \citep{lorkowski2012}. The resolved model region is shown in \figref{fig:ModelArea}. The model applies a horizontal resolution of 1/5\,$^\circ$ with 82 grid points in latitudinal direction and 1/3\,$^\circ$ with 88 grid points in longitudinal direction. The vertical dimension is resolved in 24 layers. The levels from the surface to the depth of 1000\,m are: 10\,m, 15\,m, 20\,m, 25\,m, 30\,m, 35\,m, 40\,m, 45\,m, 50\,m, 60\,m, 75\,m, 100\,m, 150\,m, 200\,m, 300\,m, 400\,m, 500\,m, 600\,m, 800\,m and 1000\,m. The time step is set to 10\,minutes.\\
\indent We focused our analysis of the new parameterisation on a station west of Ireland in the eastern Rockall Trough (55$^\circ$\,3'\,24''\,N, 10$^\circ$\,14'\,W, see marked position in \figref{fig:ModelArea}). This station was selected because winter convection reaching depths to more than 500\,m has regularly been reported in the Rockall Trough \citep{holliday2000}. \citet{meincke1986} reported MLD of up to 1000\,m during severe winters. Furthermore, it is one of the few stations where chlorophyll measurements are available for the winter season. The observation data was provided by the British Oceanographic Data Centre (BODC) and was measured on February 27th, 1996, during the cruise CH125B of RRS Challenger \citep{hill1996}.\\
\indent The hydrodynamical model HAMSOM was initialised with a monthly-averaged climatology based on the World Ocean Atlas \citep[WOA;][]{conkright2002}. Salinity was treated as a semi-prognostic variable and adjusted to the climatology with a time constant of 14 days to guarantee reasonable salinity distributions. At the open boundaries temperature and salinity are prescribed during inflow situations with a time constant of 7 days. Additionally, the surface elevation according to the M2 tide was prescribed at the open boundaries. The meteorological forcing including air temperature, cloud coverage, relative humidity, wind speed and direction was calculated from NCEP/NCAR reanalysis data \citep{kalnay1996} and was provided as 6-hourly values. Starting from the initial data, the HAMSOM model applied a 10 day spin-up during which the air pressure was constant and the wind speed was zero. The results of the hydrodynamical simulation for 1996 were calculated as averages over two tidal cycles representing daily values. This simulation output served as the basis for the biogeochemical simulation with ECOHAM.\\
\indent To demonstrate the effect of the new parameterisation, we compared a simulation run using the new parameterisation (hereafter `phytoconvection run') to a simulation using the conventional parameterisation (hereafter `standard run'), using equation \eqref{equ:PB_total} and equation \eqref{equ:PB_Steele}, respectively. For both simulations, the initialisation of the biogeochemical state variables was done using a dataset produced by a model run using only the conventional parameterisation following equation \eqref{equ:PB_Steele}. All other settings were according to \citet{lorkowski2012}.
\subsection{The convection model} \label{subsec:ConvMod}
\indent A non-hydrostatic convection model was used to evaluate the assumption that the MLD is persistently mixed by convection, and thus, constitutes a valid indicator for the vertical extent of the convective cell. It was furthermore used to test the validity of the ratio of $t_{exp}/t_{orb}$. The applied convection model uses the Boussinesq equations for an incompressible fluid in a 2,5-dimensional ocean slice with cycling boundary conditions, thus, allowing to include the rotational effects of the earth. For the turbulent eddy viscosity the turbulence closure scheme by \citet{kochergin1987} is used. The model calculates the density (temperature and salinity), the hydrodynamic flow fields and the non-hydrostatic pressure over space and time. It neglects the influence of wind stresses but allows latent and sensible heat fluxes due to fluctuations in the wind speed. Thus, a free convective boundary layer is assumed. For a detailed description of the model, including the equations, the reader is referred to \citet{kaempf1998} and to \citet{wehde2000}.\\
\indent The model was set up on a domain of 1000\,m depth and 250\,m width with a horizontal and vertical grid size of 5\,m applying a time step of 10\,seconds. We conducted two simulations starting at March 17th and March 29th to capture the transition period between winter convective mixing and the decline of the mixed layer, respectively. Each simulation ran over a period of 14 days with one additional day as spin-up. The initial profiles of temperature and salinity for the first simulation were vertically interpolated from the simulated profiles of the 3D simulation at the station in the Rockall Trough (see \figref{fig:ModelArea}). The second simulation was initialised with the results of the first one. The same meteorological forcing as for the 3D simulation for the respective station was used, provided as 3-hourly interpolated data. To test the validity of the use of the ratio of $t_{exp}/t_{orb}$ 200 Lagrangian tracers were randomly distributed within the mixed layer at the beginning of the first simulation. They record the actual light throughout the simulation period, thus, allowing the calculation of $t_{exp}$ and $t_{orb}$ for each tracer.
\section{Results and Discussion} \label{sec:Results&Discussion}
\subsection{The three-dimensional physical-biogeochemical model} \label{subsec:Res3DMod}
\indent We will focus our discussion on the biogeochemical simulations as the physical simulation only built the basis for testing the parameterisation of phytoconvection. For this purpose, only a brief presentation of the simulated temperature and the derived MLD is given. All Hovm{\"o}ller diagrams, time series and the comparison to the observations refer to the station in the Rockall Trough marked in \figref{fig:ModelArea}.
\subsubsection{Hydrodynamic simulation} \label{subsubsec:ResHydro}
\begin{figure}[H]
\centering
\includegraphics[width=0.6\textwidth]{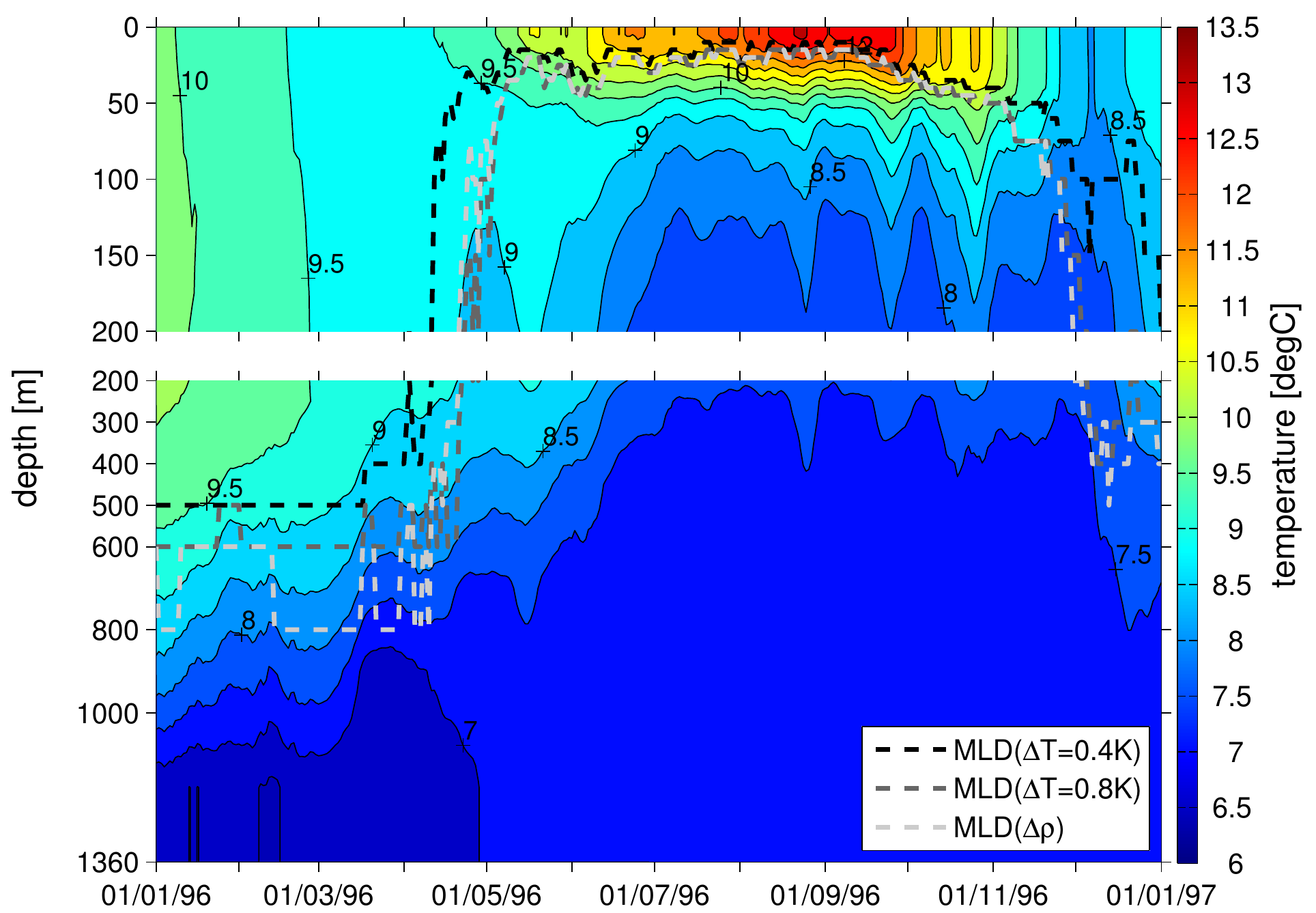}
\caption{Hovm{\"o}ller diagram of simulated temperature for 1996. The dashed lines depicts the MLD determined according to different MLD criteria: the black line refers to MLD after equation \eqref{equ:CML_crit}, the dark grey and light grey lines refer to the MLD determined with a temperature difference and, respectively, density difference criterion based on $\Delta T = 0.8$\,K \citep{kara2000}.}
\label{fig:TempHov}
\end{figure}

\indent The temporal development of the simulated temperature at the station in the Rockall Trough (see \figref{fig:ModelArea}) and the corresponding MLD (dashed lines) are shown in \figref{fig:TempHov}. The MLD resulting from equation \eqref{equ:CML_crit} (black line) is compared to the MLD determined using a temperature difference (dark grey line) and density difference criterion (light grey line) after \citet{kara2000} based on a $\Delta T$ of 0.8\,K. The simulation starts from the climatological dataset with mixed layer temperatures of above 10\,$^\circ$C which steadily decrease until mid-March due to ongoing surface cooling and deep mixing. During this period the MLD determined after equation \eqref{equ:CML_crit} is at maximum 500\,m which is in the same order as the climatological MLD reported for the northern Rockall Trough at about 57\,$^\circ$N to 58\,$^\circ$N \citep{holliday2000}, which is at maximum about 700\,m deep. In April the mixed layer declines and a shallow seasonal surface mixed layer develops reaching maximum temperatures of above 13.5\,$^\circ$C in August. Starting in October, the MLD deepens again reaching depths of up to 200\,m at the end of the year. Throughout the whole simulation period the two different MLDs determined after \citet{kara2000} show greater values than the applied criterion. From January to March the two criteria yield MLDs 100\,m to 300\,m deeper than the MLD criterion used in our study, which contradict the temperatures visually decreasing already in 300\,m to 400\,m depth. In mid-April the strong decrease in the applied MLD coincides with the beginning of near-surface thermal stratification whereas the two MLDs after \citet{kara2000} stay deep until late April showing. This supports the use of the applied MLD criterion.\\
\begin{figure}[H]
\centering
\subfigure{
\includegraphics[width=0.475\textwidth]{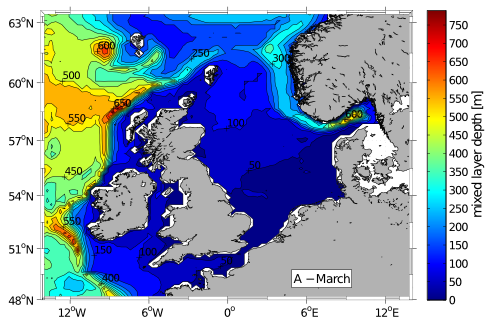}
}
\subfigure{
\includegraphics[width=0.475\textwidth]{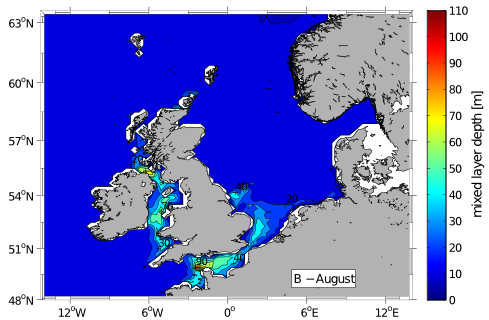}
}
\caption{Monthly averaged simulated MLD within the model region determined according to $SST - T = 0.4$\,K (see equation \eqref{equ:CML_crit}) for March (A) and April (B) 1996. The different colour scales should be noticed.}
\label{fig:MLDSurf}
\end{figure}
\indent The monthly averaged MLD after \eqref{equ:CML_crit} for March and August within the whole model area is shown in \figref{fig:MLDSurf}. The average MLD of 400\,m to 600\,m in March in the Rockall Trough region are in good agreement with \citet{holliday2000}. In most parts of the shelf the MLD represents the bottom topography due to strong winter mixing reaching down to the bottom. In August most parts of the model area are stratified with MLD less than 20\,m representing a small underestimation \citep{paetsch2008}. Only parts of the English Channel, the Irish Sea and the southern North Sea show a MLD reaching the bottom. This is in good agreement with \citet{pingree1978b} who reported on perennial well-mixed waters in these regions due to strong tidal mixing.\\
\subsubsection{Biogeochemical simulations} \label{subsubsec:ResBiogeo}
\begin{figure}[H]
\centering
\includegraphics[width=0.5\textwidth]{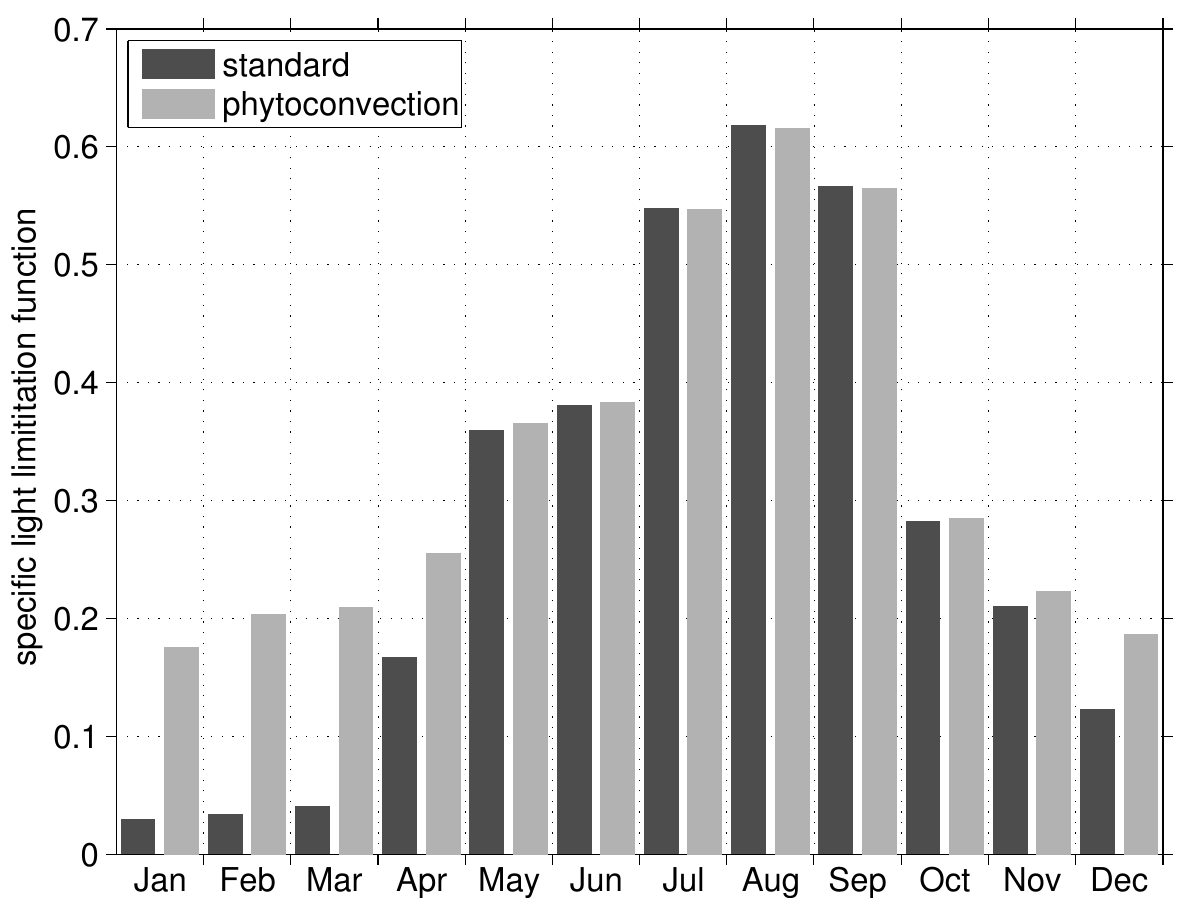}
\caption{Time series of the monthly and vertically averaged specific light limitation function within the mixed layer for the standard run (dark grey) and the phytoconvection run (light grey). The specific light limitation function is defined as the light-dependent production rate (see equation \eqref{equ:PB_total}) normalised by the maximum production rate $P^B_{max}$.}
\label{fig:LightLim}
\end{figure}

\begin{figure}[H]
\centering
\includegraphics[width=0.6\textwidth]{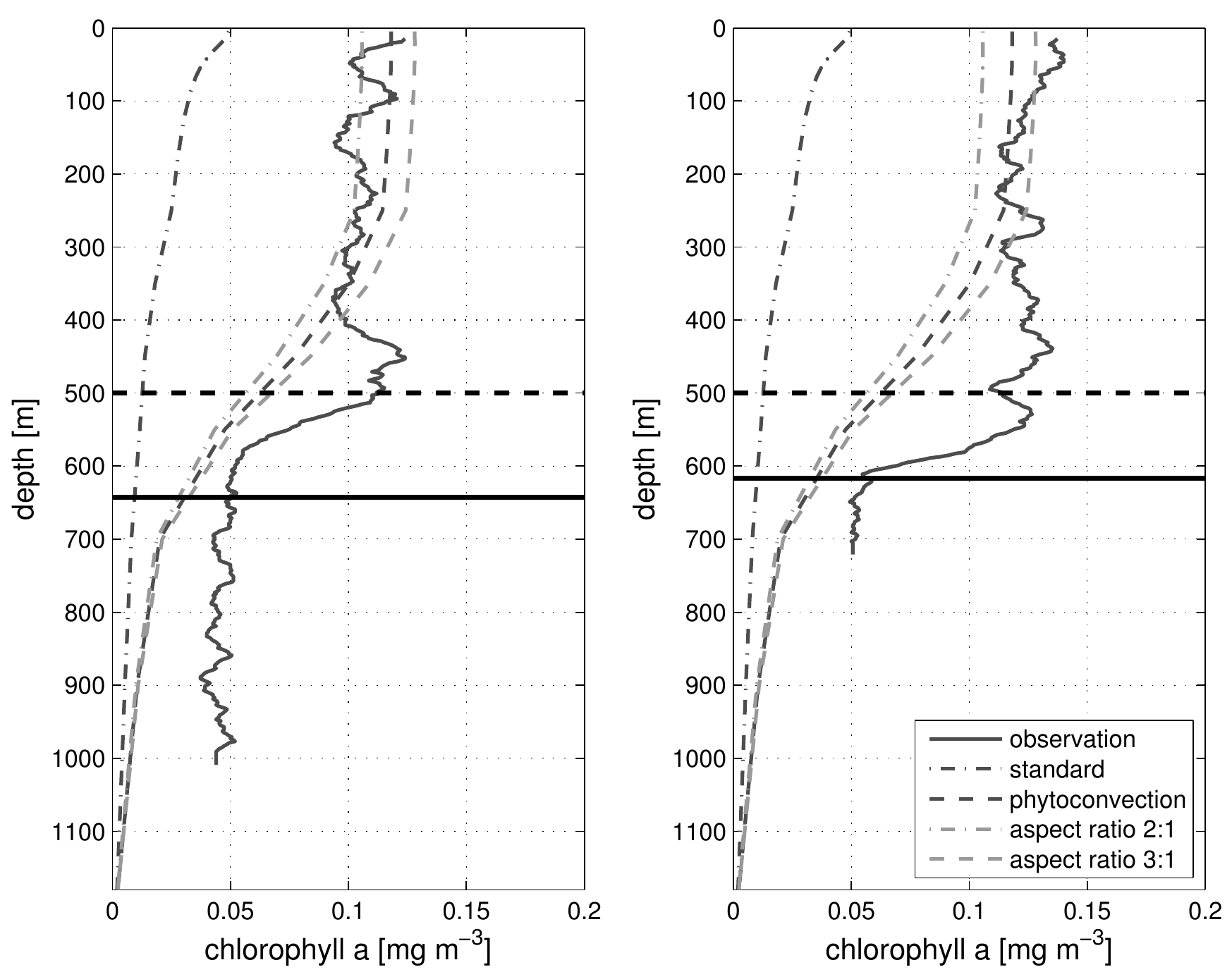}
\caption{Comparison of observed (solid) and simulated (dashed/dash-dotted) chlorophyll-a profiles on February 27th, 1996. Simulations presented: standard run (dark grey, dash-dotted), phytoconvection run with aspect ratio of 2.5:1 (dark grey, dashed), phytoconvection run with aspect ratio of 2:1 (light grey, dash-dotted) and phytoconvection run with aspect ratio of 3:1 (light grey, dashed). Horizontal black lines depict the MLD referring to the observations (solid) and the simulations (dashed).}
\label{fig:Chla_OBSvsSIM}
\end{figure}

\begin{figure}[H]
\centering
\subfigure{
\includegraphics[width=0.475\textwidth]{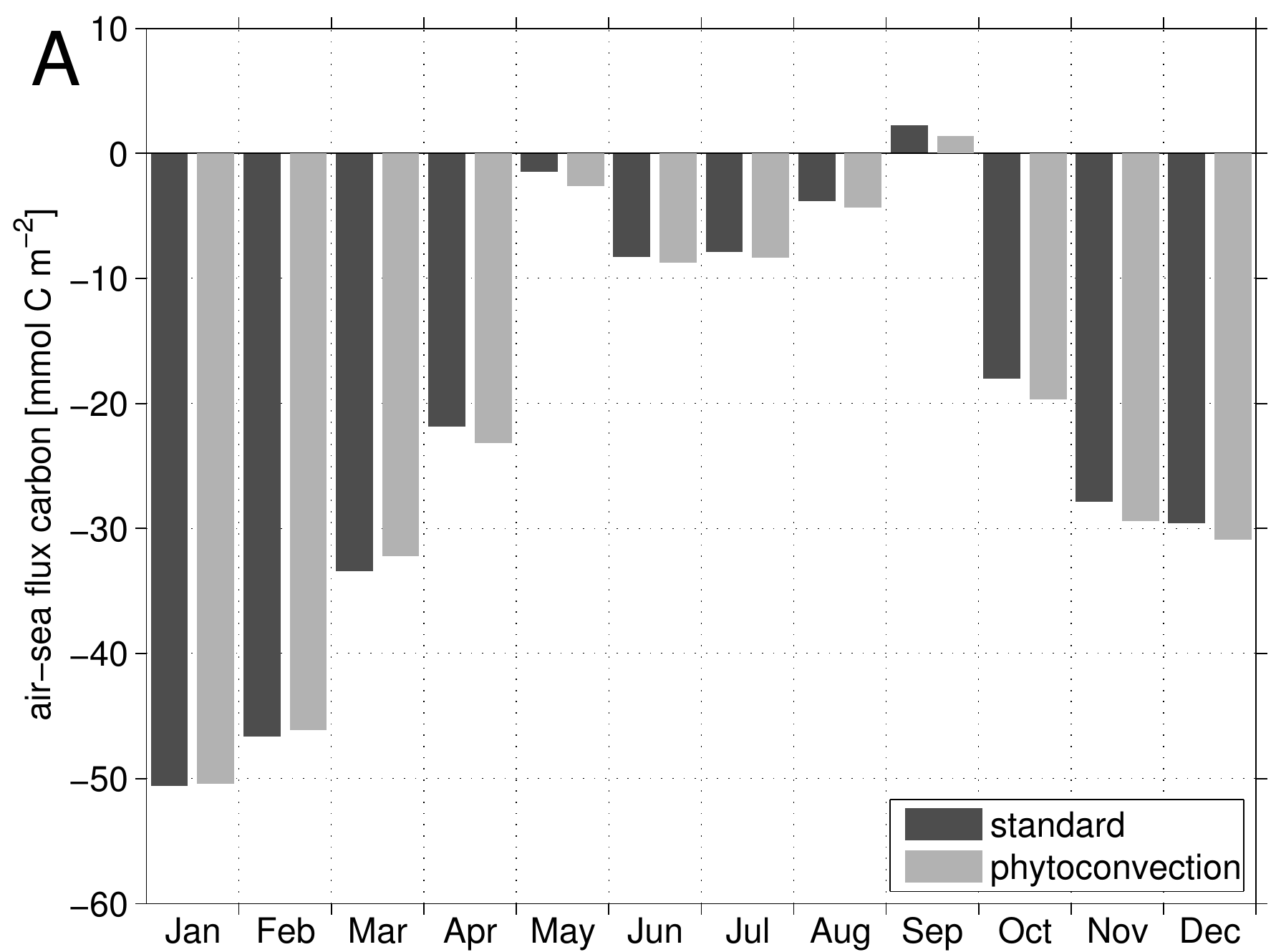}
}
\subfigure{
\includegraphics[width=0.475\textwidth]{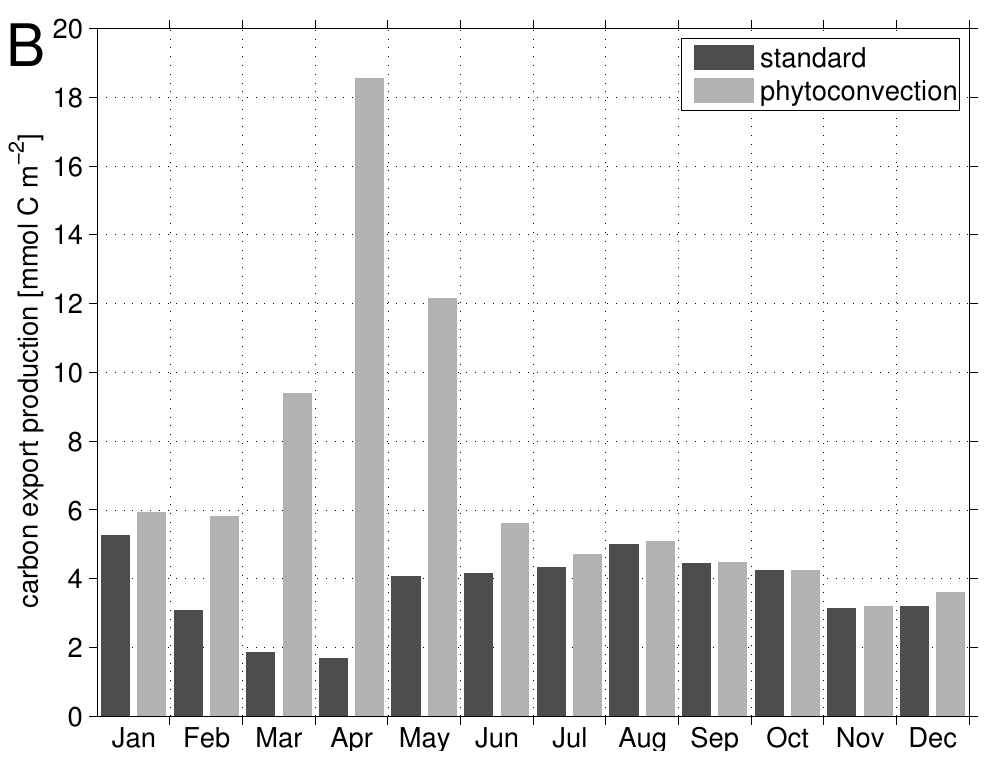}
}
\caption{Time series of monthly integrated (A) air-sea carbon flux and (B) carbon export production for the standard run (dark grey) and the phytoconvection run (light grey). The negative values in the air-sea flux mean outgassing of CO$_2$. The carbon export production is defined as the export of fast-sinking detritus below 500\,m depth referring to the maximum MLD at the analysed station.}
\label{fig:AirSea_CEP}
\end{figure}

\indent \figref{fig:LightLim} shows the monthly and vertically averaged specific light limitation function (LLF) within the mixed layer for the standard run (dark grey) and the phytoconvection run (light grey) at the station west of Ireland (see \figref{fig:ModelArea}). The LLF is defined as the light-dependent production rate (equation \eqref{equ:PB_total}) normalised by the maximum production rate $P^B_{max}$, and therefore, is a direct measure for the phytoplankton production.\\
\indent The two simulations generally follow the seasonal cycle of the solar radiation showing low values during winter and high values during summer. Because of the maximum influence of the phytoconvective parameterisation, the phytoconvection run shows significantly higher values than the standard run during winter due to the higher values of the LLF in the deeper part of the mixed layer.\\
\begin{figure}[H]
\centering
\subfigure{
\includegraphics[width=0.6\textwidth]{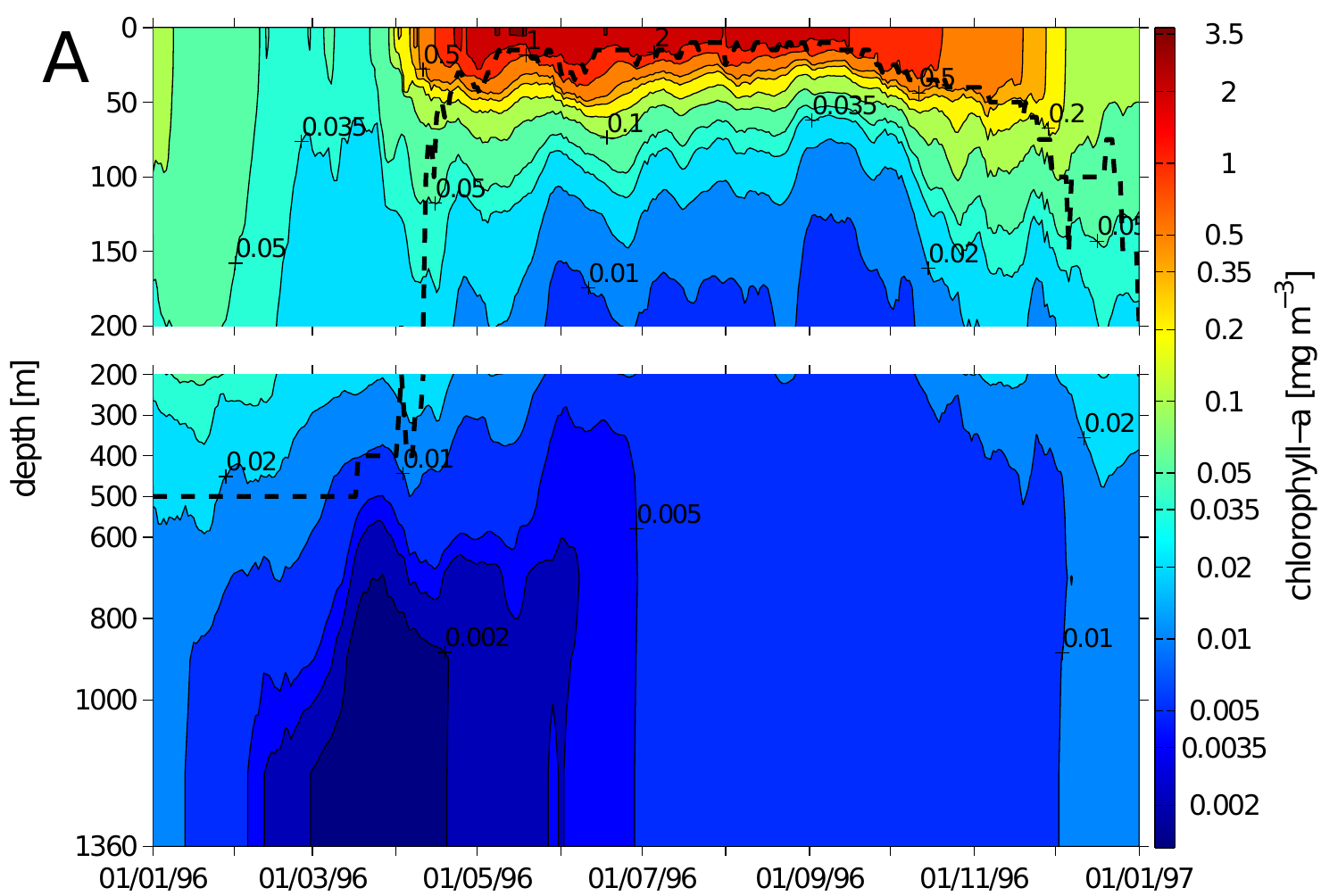}
}
\subfigure{
\includegraphics[width=0.6\textwidth]{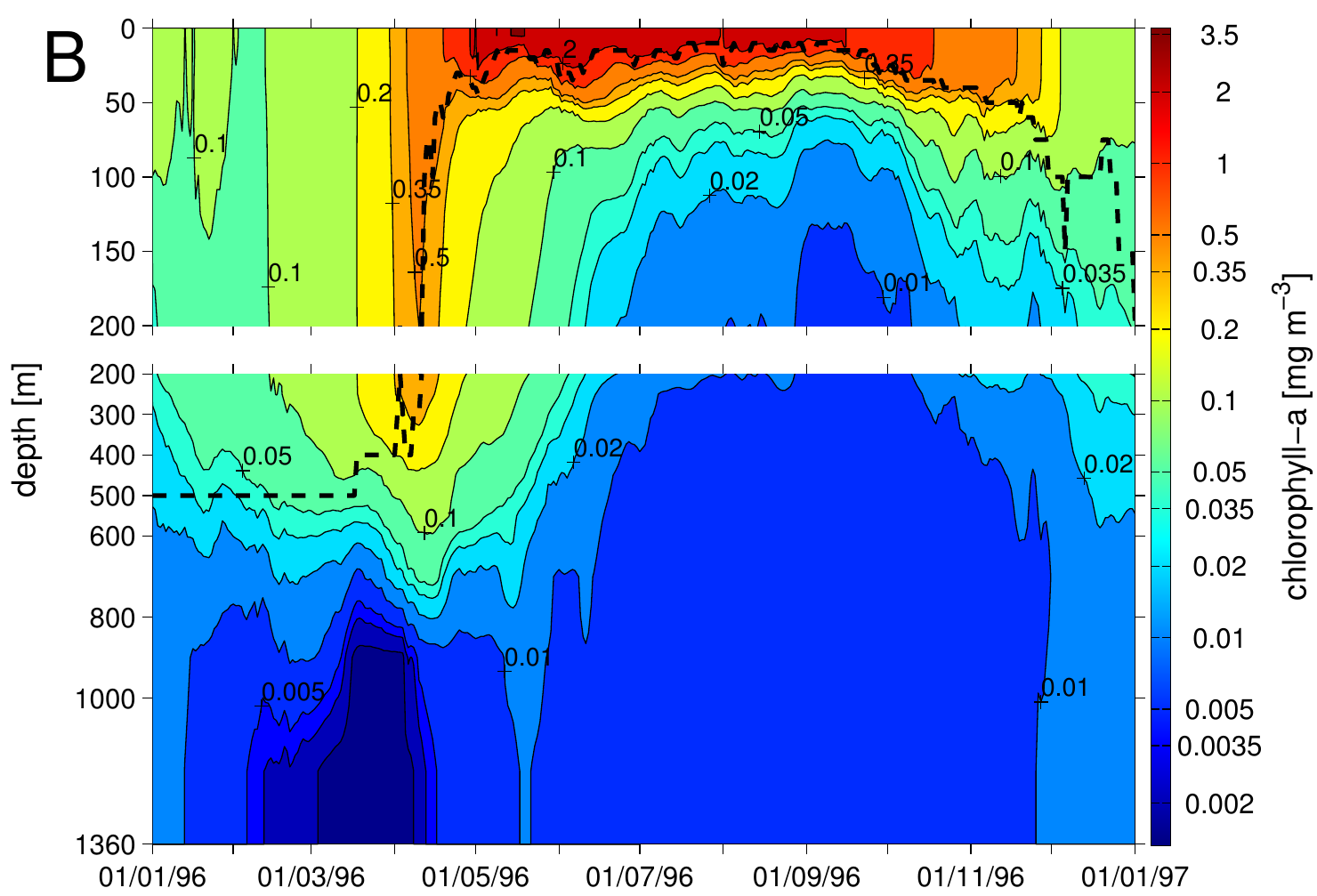}
}
\caption{Hovm{\"o}ller diagrams of simulated chlorophyll-a concentrations for (A) the standard run and (B) the phytoconvection run. The chlorophyll-a values were derived from simulated phytoplankton carbon applying a constant chl-a:C mass ratio of 1:50. The dashed lines depict the MLD. The logarithmic colour scales should be noticed.}
\label{fig:ChlaHov}
\end{figure}
\indent From May to October the two simulations show similar LLFs, due to the shallow seasonal mixed layer causing the influence of phytoconvection to be zero most times. Small differences between the two simulations during this period are caused by the interaction of the conventional parameterisation and the phytoconvective parameterisation during events of wind-induced mixed layer deepening. At the end of the year the LLFs are again diverging due to the deepening of the mixed layer, and thus, the increasing influence of phytoconvection.\\
\indent \figref{fig:ChlaHov} shows the Hovm{\"o}ller diagrams of chlorophyll-a (chl-a) simulated by the standard run (A) and the phytoconvection run (B) at the station west of Ireland. The chl-a concentrations were derived from the simulated phytoplankton carbon applying a constant chl-a:C mass ratio of 1:50. The phytoplankton carbon is the sum of carbon of the two simulated phytoplankton groups diatoms and flagellates. The two simulations start from the same initial conditions with highest concentrations of above 0.1\,mg\,chl-a\,m$^{-3}$ in the upper 100\,m and decreasing concentrations in greater depths.\\
\indent In the first two weeks, the two simulations show decreasing concentrations in the upper 100\,m and, due to downward mixing of the higher surface concentrations, increasing concentrations in the layers in between 100\,m and the MLD. Thereafter, the chl-a of the two simulations diverge. The standard run shows a steady decrease until late March throughout the whole water column while the phytoconvection run is characterised by steadily increasing concentrations throughout the mixed layer from early February until mid-April. This development is controlled by the different LLFs \figref{fig:LightLim}. The concentrations in the standard run drop below 0.05\,mg\,chl-a\,m$^{-3}$ in the upper 100\,m and values less than 0.035\,mg\,chl-a\,m$^{-3}$ in the deeper layers, while the phytoconvection run shows concentrations of above 0.05\,mg\,chl-a\,m$^{-3}$ throughout the whole mixed layer and concentrations higher than 0.1\,mg\,chl-a\,m$^{-3}$ in the upper 200\,m to 300\,m. Maximum concentrations in the deep part of the winter mixed layer are reached in early April when the MLD declines and a seasonal thermocline develops with values of above 0.35\,mg\,chl-a\,m$^{-3}$ in the phytoconvection run. At the same time the standard run shows the onset of a strong surface spring bloom indicated by the increase in the chl-a concentrations from less than 0.05\,mg\,chl-a\,m$^{-3}$ to above 1\,mg\,chl-a\,m$^{-3}$ within about two weeks. A comparably strong surface bloom is not simulated in the phytoconvection run, because the MLD is still 200\,m to 400\,m deep, and hence, the phytoconvective parameterisation is fully taken into account.\\
\indent From late April to November the two simulations show a very similar behaviour within the mixed layer as expected from the similarity of the LLFs during this period (see \figref{fig:LightLim}). Caused by the increasing influence of the phytoconvective parameterisation in December, the phytoconvection run shows slightly higher concentrations than the standard run in the deeper layers (50\,m to 300\,m).\\
\indent \figref{fig:ChlaHov} showed that the phytoconvection run simulates significantly different winter chl-a concentrations than the standard run. In order to evaluate the simulated chl-a profiles, the standard run (dark grey, dash-dotted) and the phytoconvection run (dark grey, dashed) are compared to two measured chl-a profiles (solid) recorded on February 27th, 1996 \citep{hill1996} (\figref{fig:Chla_OBSvsSIM}). The horizontal black lines mark the MLD determined using equation \eqref{equ:CML_crit} from the simulated (dashed) and measured (solid) temperature profiles, respectively. The two observed profiles show a distinct structure with increased chl-a concentrations ranging between 0.09\,mg\,chl-a\,m$^{-3}$ and 0.14\,mg\,chl-a\,m$^{-3}$ in the upper 500\,m to 600\,m. Thereunder, concentrations strongly decrease followed by concentrations of about 0.05\,mg\,chl-a\,m$^{-3}$ below the MLD. Chl-a concentrations in the upper hundreds of metres are in the same order of magnitude as measurements made in different regions of the North Atlantic \citep{backhaus2003}.\\
\indent The standard run shows maximum concentrations of about 0.05\,mg\,chl-a\,m$^{-3}$ at the surface and steadily decreasing concentrations with increasing depth. In contrast, the phytoconvection run shows increased concentrations of about 0.11\,mg\,chl-a\,m$^{-3}$ to 0.12\,mg\,chl-a\,m$^{-3}$ in the upper 200\,m to 300\,m followed by a stronger decrease and concentrations less than 0.02\,mg\,chl-a\,m$^{-3}$ below 700\,m. In the upper 200\,m to 300\,m these results are comparable with the observations, even though the vertical variability of the measurements is not reproduced. The standard run shows significantly lower chl-a concentrations throughout the whole water column and does not reflect the distinct vertical structure of the observations. The vertical structure of the phytoconvection run fits the observations much better with its higher concentrations in the upper hundreds of metres and the subsequent stronger gradient between 300\,m and 700\,m with the maximum gradient around the MLD. Nevertheless, the simulated chl-a is not homogeneously distributed throughout the whole mixed layer which may be due to the vertical mixing of the phytoplankton to deeper layers.\\
\indent In the observations the determined MLD is deeper than the depth of the maximum gradient in the chl-a which indicates that the applied MLD criterion may overestimate the actual MLD. In the simulation the maximum gradient corresponds to the simulated MLD. However, it remains constrained by the vertical resolution of the model grid which is only 100\,m between 200\,m and 600\,m, and therefore is not able to reflect the sharp gradient of the observations.\\
\indent This comparison to observations is rather basic due to the low data availability and gives just a first insight in the capabilities of the new parameterisation. Hence, a more compelling validation is required for a complete evaluation of the presented parameterisation, which we leave to future work.\\
\indent The comparison of the chl-a concentrations with those in \citet{levy1998a} in February (the month with maximum MLD in their study) shows that our parameterisation yields concentrations in the same order of magnitude (about 0.1\,mg\,chl-a\,m$^{-3}$) while chl-a concentrations at the beginning of the period simulated by \citet{levy1998a} are about twice as high compared to our simulation. Thus, the parameterisation by \citet{levy1998a,levy1998b} applied on our initial conditions wouldmost likely result in chl-a concentrations underestimating the observations by about factor 2. Furthermore, the approach presented here is reversed to the approach by \citet{levy1998a,levy1998b} regarding the effect of convection on primary production. In our parameterisation primary production is increased by convection while \citet{levy1998a,levy1998b} included convective mixing as a limiting factor for primary production. The parameterisation by \citet{levy1998a,levy1998b} applies an additional factor $\gamma_m$ ranging between 0.1 and 1 representing maximum limitation of growth due to vertical mixing in the case of $H_{cml} > 2\cdot H_{euph}$ and no convection-induced limitation for $H_{cml} < H_{euph}$, respectively. However, this parameterisation has not been applied to mixed layers deeper than 200\,m and \citet{levy1998a} do not deal with cases of $H_{cml} > 2\cdot H_{euph}$. Thus, the winter chl-a concentrations simulated by \citet{levy1998a} would be reduced further in the case of a winter MLD of similar depth as in our simulations. Consequently, the parameterisation by \citet{levy1998a,levy1998b} is unlikely to reproduce the observed winter chl-a concentrations. We use the vertically averaged LLF which is subsequently distributed over the MLD whereas \citet{levy1998a,levy1998b} used the vertically averaged PAR. The advantage of using the vertically averaged LLF is, that it allows for photoinhibition which might occur in the surface layer when solar radiation increases in early spring. In contrast, using the average PAR reduces the PAR in the upper layers, and thus, can prevent photoinhibition resulting in an overestimation of growth.\\
\indent The analysis of the phytoconvection run with respect to nutrients (specifically nitrate, not presented here) showed that even though the increased primary production during winter affects the vertical nutrient distribution, the levels stay well above any limiting thresholds. For the zooplankton grazing (also not presented) there are some differences between the standard run and the phytoconvection run, especially in the period from February to April. The monthly and vertically integrated grazing rates in the phytoconvection run exceed those in the standard run by above factor 8 during March. However, during the bloom initiation in mid-April the daily zooplankton grazing (not shown here) in the surface layer is very similar in both simulations (maximum deviations of about 13\% with even higher rates in the standard run), indicating only a small impact on the phytoplankton development.\\
\indent In ECOHAM, zooplankton grazing is directly proportional to the phytoplankton concentration which explains the stronger grazing in the phytoconvection run during winter. The model implicitly accounts for the decoupling of zooplankton and phytoplankton as zooplankton grazing declines with decreasing phytoplankton concentrations, and consequently zooplankton concentrations decrease. These dynamics are in agreement with the dilution phase of the 'Dilution-Recoupling-Hypothesis' \citep{behrenfeld2010}, representing one possible approach for how winter phytoplankton concentrations can be reproduced in models. The 'recoupling phase' \citep{evans1985,behrenfeld2010} however is not captured by the model, since zooplankton, forced by the vertical exchange coefficient $A_V$ (see equation \eqref{equ:VertExchange}), is mixed within the water column like other state variables, and thus, is unable to retain its position within the water column.\\
\begin{figure}[H]
\centering
\subfigure{
\includegraphics[width=0.475\textwidth]{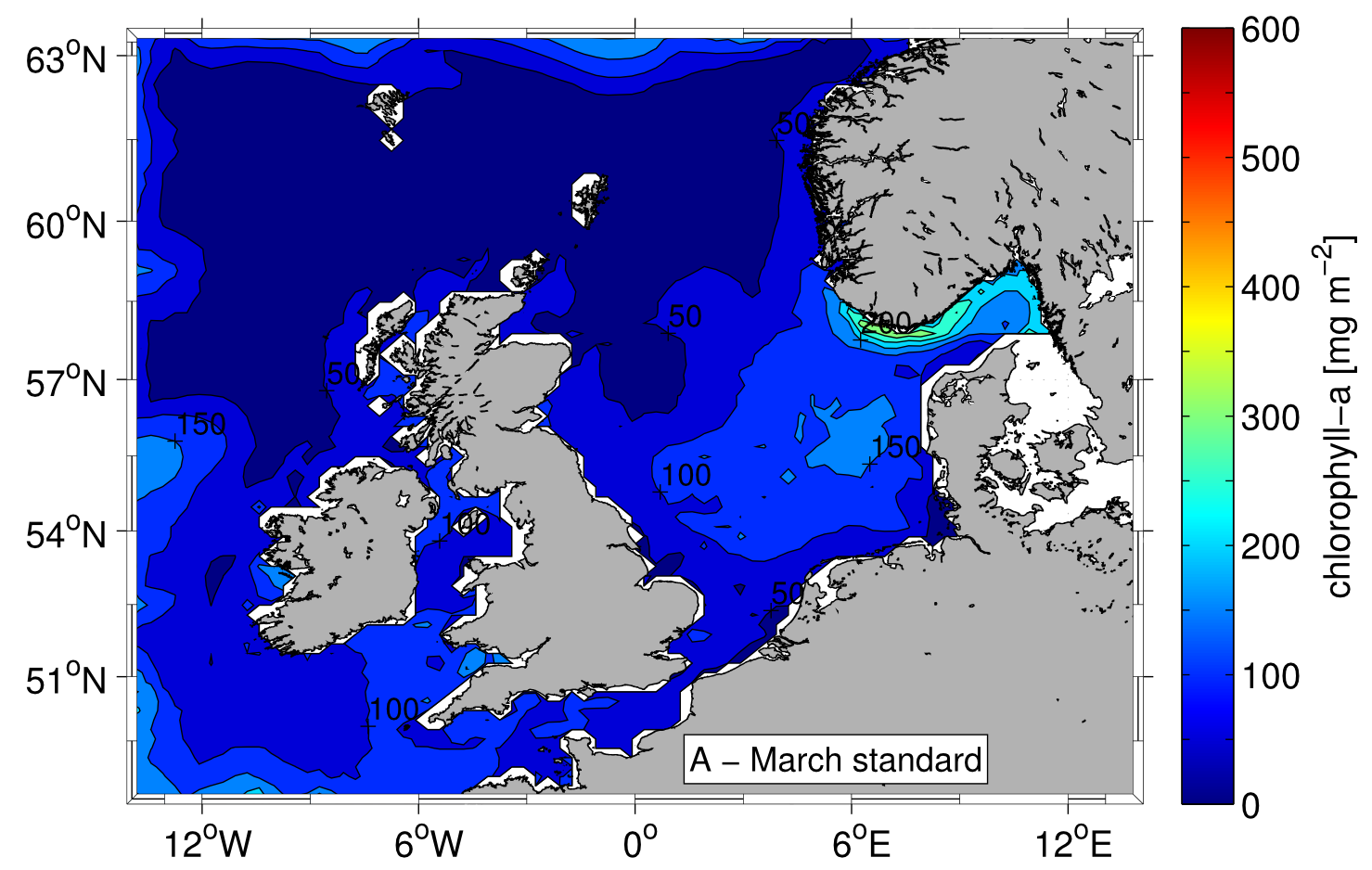}
}
\subfigure{
\includegraphics[width=0.475\textwidth]{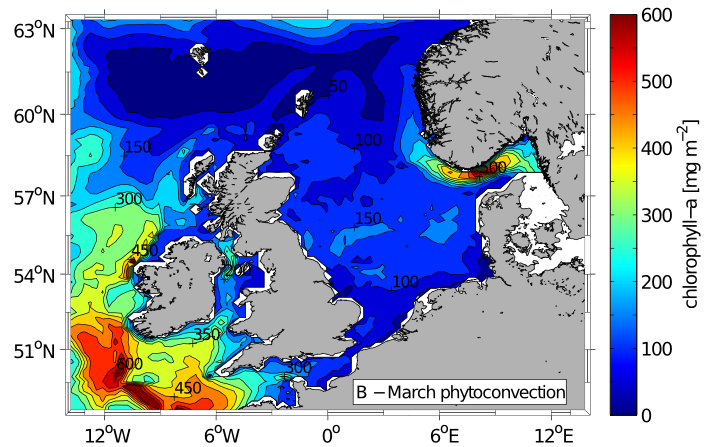}
}
\subfigure{
\includegraphics[width=0.475\textwidth]{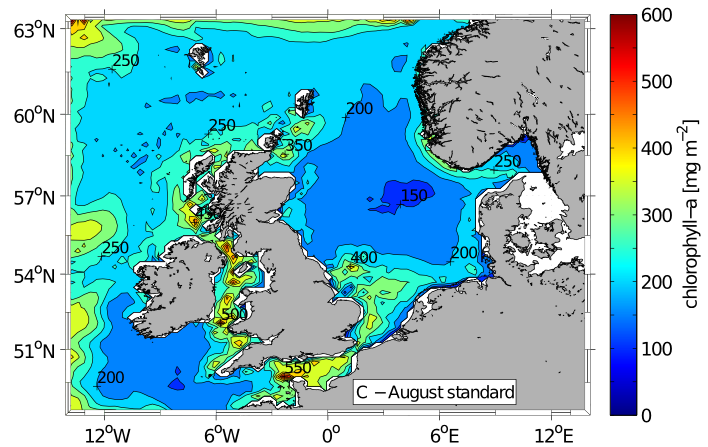}
}
\subfigure{
\includegraphics[width=0.475\textwidth]{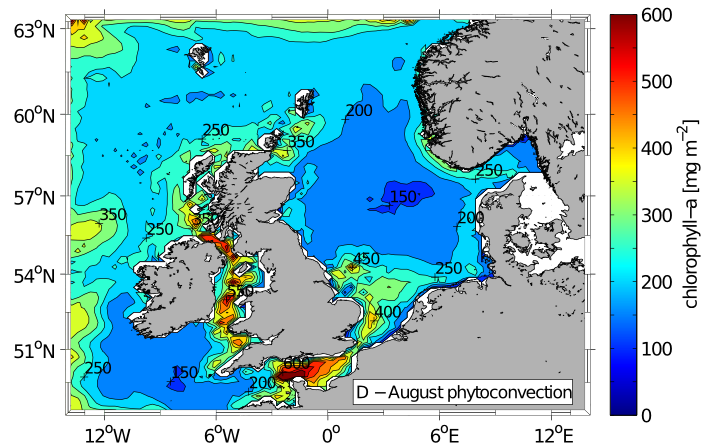}
}
\caption{Monthly averages of vertically integrated chlorophyll-a in the model region for (A) the standard run in March, (B) the phytoconvection run in March, (C) the standard run in August and (D) the phytoconvection run in August. The integration depth is 500\,m.}
\label{fig:ChlaSurf}
\end{figure}
\indent To analyse the functioning of the developed weighting function \eqref{equ:f_p} \figref{fig:ChlaSurf} shows the vertically integrated, monthly averaged chl-a concentrations of the upper 500\,m for March (A and B) and August (C and D). In March the standard run (A) shows significantly lower concentrations than the phytoconvection run (B) in the areas south and west of Ireland due to the MLD being deeper than the reference depth $H_{ref} = 100$\,m (see \figref{fig:MLDSurf}). The highest concentrations in these regions exceed 600\,mg\,chl-a\,m$^{-2}$ in the phytoconvection run while the concentrations in the standard run stay below 150\,mg\,chl-a\,m$^{-2}$ in the same area.\\
\indent In the deep areas north of the shelf the concentrations in both simulations are similarly low despite the deep mixed layers. Here, phytoplankton concentrations are very low at the beginning of the simulation (about factor 20 less than at the analysed station) and stay low throughout the winter. South of Norway both simulations show increased chl-a concentrations, but with a spatial mismatch. Due to the deep mixed layer the phytoconvection run produces the highest concentrations in the deepest area of the Skagerrak. In contrast, the standard run shows the highest concentrations directly at the coast due to the onset of the spring bloom. In the shallower shelf regions, e.g. the southeastern North Sea the two simulations are in good agreement with each other demonstrating that the weighting function allows the application of the new parameterisation on the shelf.\\
\indent In August (C and D) the two simulations show the same patterns and concentrations in most parts of the shelf and the areas off the shelf. The shallow seasonal mixed layer in these areas (see \figref{fig:MLDSurf}) causes phytoconvection in the phytoconvection run (D) to switch off. Only in the English Channel and the Irish Sea the phytoconvection run shows significantly higher concentrations than the standard run (C). In these regions the bottom depth is around 100\,m and the vertical mixing is strong throughout the whole year due to tidal mixing and the interaction of horizontal currents with the bottom topography. This prevents the development of a persistent seasonal mixed layer, i.e. the MLD deepens frequently causing the influence of phytoconvection to increase. Consequently, the chl-a concentrations in the deeper layers are significantly higher than in the standard run resulting in higher vertically integrated concentrations. These regions are also known for the regular occurrence of tidal fronts and increased phytoplankton production during summer due to the availability of light and nutrients \citep{pingree1978a}. This indicates reasonable chl-a concentrations simulated in the phytoconvection run in these regions.\\
\indent The monthly and vertically integrated (0\,m to 500\,m) net primary production (NPP) in the whole model region shows similar patterns as the chl-a (\figref{fig:ChlaSurf}) (not shown). In regions where the influence of phytoconvection is strongest (southwest of Ireland) the NPP in the phytoconvection run exceeds the NPP in the standard run by about factor 8 to 10 during March. Maximum values of above 18\,g\,C\,m$^{-2}$\,month$^{-1}$ are reached in these areas in the phytoconvection run. In August, the phytoconvection run reaches maximum NPP of above 30\,g\,C\,m$^{-2}$\,month$^{-1}$ in the English Channel while a maximum NPP of about 16\,g\,C\,m$^{-2}$\,month$^{-1}$ is reached south of the Doggerbank in the standard run.\\
\indent To illustrate the effect of the new parameterisation on the carbon cycle \figref{fig:AirSea_CEP} shows the monthly integrated air-sea carbon flux (ASCF, A) and the carbon export production (CEP, B). Negative values in the ASCF imply outgassing of CO$_2$. The CEP is defined as the export of fast-sinking detritus below 500\,m depth as this is the maximum MLD at the specific station. For the ASCF the two simulations are only slightly different from each other throughout the whole seasonal cycle. During winter the two simulations show maximum outgassing of CO$_2$, or the most negative ASCF due to strong mixing, and hence, the transport of carbon-enriched deep water to the surface. During this time the phytoconvection run (light grey) shows slightly higher values induced by the higher phytoplankton biomass, and therefore, increased primary production at the surface compared to the standard run (dark grey).\\
\indent During the spring bloom formation, the ASCF increases significantly due to the increasing near-surface primary production. As near-surface primary production during April and May is stronger in the standard run than in the phytoconvection run, the relation between the two simulations shifts and the ASCF becomes higher in the standard run. During summer the ASCF stays on a relatively high level due to the ongoing primary production. The slightly lower values in the phytoconvection run are caused by wind-induced mixed layer deepening, and thus, the influence of phytoconvection which lowers the primary production in the surface layer. From October to December outgassing again intensifies due to increased mixing which brings carbon-enriched water to the surface, and the reduced primary production. The standard run shows only slightly higher values. This is caused by the increased importance of the phytoconvection leading to lower near-surface production rates, while the near-surface phytoplankton biomass remains similar in the two simulations after summer (see \figref{fig:ChlaHov}).\\
\indent While the ASCF is very similar for the two simulations throughout the year, the CEP shows significant differences during the annual cycle. In the standard run the CEP is steadily decreasing from January to April when the spring bloom is fully developed due to the decreasing phytoplankton biomass. In contrast, the phytoconvection run shows a steady increase in the CEP from February to April with maximum amounts being 6 to 9 times higher than in the standard run. This is caused by the significantly higher biomass of phytoplankton and zooplankton in the water column from the surface to the MLD, and consequently, higher amounts of fast-sinking detritus at great depths. The peak in April with a CEP of about 18.5\,mmol\,C\,m$^{-2}$ is related to the strong decline of the mixed layer leaving a large portion of the phytoplankton biomass below the mixed layer which is than transferred to detritus sinking to the deep ocean. During summer the two simulations are again converging and show a similar CEP until October. Thereafter, the CEP becomes again slightly higher in the phytoconvection run due to the increasing influence of phytoconvection, and hence, the slightly higher concentrations of phytoplankton in greater depths (see \figref{fig:ChlaHov}).\\
\indent As the spatial aspect ratio of the convection cell (see \figref{fig:conv_cell}) is a key factor for the parameterisation of phytoconvection we compared the chl-a simulated by the use of three different aspect ratios. \figref{fig:Chla_OBSvsSIM} shows the effect of the applied spatial aspect ratios of the convection cell on the phytoplankton. The previously discussed phytoconvection run (dark grey, dashed) applying an aspect ratio of 2.5:1 is compared to simulations applying aspect ratios of 2:1 (light grey, dash-dotted) and 3:1 (light grey, dash-dotted), respectively. The presented ratios cover the range reported by \citet{kaempf1998}. Simulated chl-a concentrations increase with rising aspect ratios as the time within the euphotic zone $t_{exp}$ increases (see equation \eqref{equ:Texp}). Furthermore, it can be seen that for the first observed profile (left panel) the 2:1 ratio produces the best fit regarding the concentrations in the upper 300\,m, while for the second observation (right panel) the 3:1 ratio fits best to the observations. This suggests, that on average the aspect ratio of 2.5:1 offers a reasonable fit.\\
\indent The deviation in the simulated chl-a of the 2:1 and 3:1 ratio simulation relative to the phytoconvection run for the purely phytoconvective period (January to March) was calculated for each time step and layer at the presented station as the Euclidean distance normalised by the average of the the two simulations. The maximum deviation for the simulation applying a ratio of 2:1 is 23.7\% reached on March 30th. The 3:1 simulation reaches its maximum of 19.8\% on March 26th. These deviations are reached when the chl-a concentrations start to significantly increase (compare \figref{fig:ChlaHov}) due to enhanced light availability. These deviations would make a more thorough validation of the results originating from different spatial aspect ratios desirable. However, this would require a more comprehensive dataset, which unfortunately was not available.\\
\indent The reference depth $H_{ref}$ controlling the weighting of the standard parameterisation and the phytoconvective parameterisation (see equations \eqref{equ:f_p} and \eqref{equ:PB_total}) was tested by two additional simulations (not shown). For the first test run we chose a deeper $H_{ref}$ of 200\,m compared to $H_{ref} = 100$\,m in the previously discussed phytoconvection run, which also enlarged the transition range, where both the standard parameterisation and the phytoconvective parameterisation are taken into account. The analysis of the monthly and vertically averaged LLF for the two cases showed that there are no differences during periods with sufficiently deep (MLD $\geq H_{ref}$) or shallow (MLD $\leq H_{euph}$) mixed layers. Whereas during the decline of the mixed layer in April and the deepening in December the relative deviation (Euclidean distance normalised by the average of the two simulations) between the two simulations is about 6\% in April and about 16.5\% in December when the MLD varies between 75\,m and 150\,m. The LLF in the simulation with $H_{ref} = 200$\,m yields a lower LLF for these periods due to the lower influence of phytoconvection, and thus, lower light availability at greater depth. This further implies that a deeper $H_{ref}$ allows for a stronger near-surface gradient in phytoplankton during the spring bloom because of the earlier reduction of the influence of phytoconvection.\\
\indent The second test run with a shallower $H_{ref}$ of 50\,m yielded a maximum increase in the LLF by about 5.7\% compared to the phytoconvection run with $H_{ref} = 100$\,m in April which is due to the stronger impact of the phytoconvective parameterisation. During the deepening of the MLD in November and December the simulation with the shallower $H_{ref}$ shows only slight differences in the LLF with maximum deviations of 3.2\% compared to the phytoconvection run which is induced by the stronger impact of the phytoconvective parameterisation. The use of a shallower $H_{ref}$ does not cause strong changes in the LLF as the transitional phase between the standard parameterisation and phytoconvection is shorter which becomes most important during the fast decline of the mixed layer in April. Conversely, a deeper $H_{ref}$ results in larger deviations in late autumn/early winter driven by the longer transitional phase during the deepening of the MLD.\\
\subsection{The convection model} \label{subsec:ResConvMod}
\begin{figure}[H]
\centering
\includegraphics[width=0.7\textwidth]{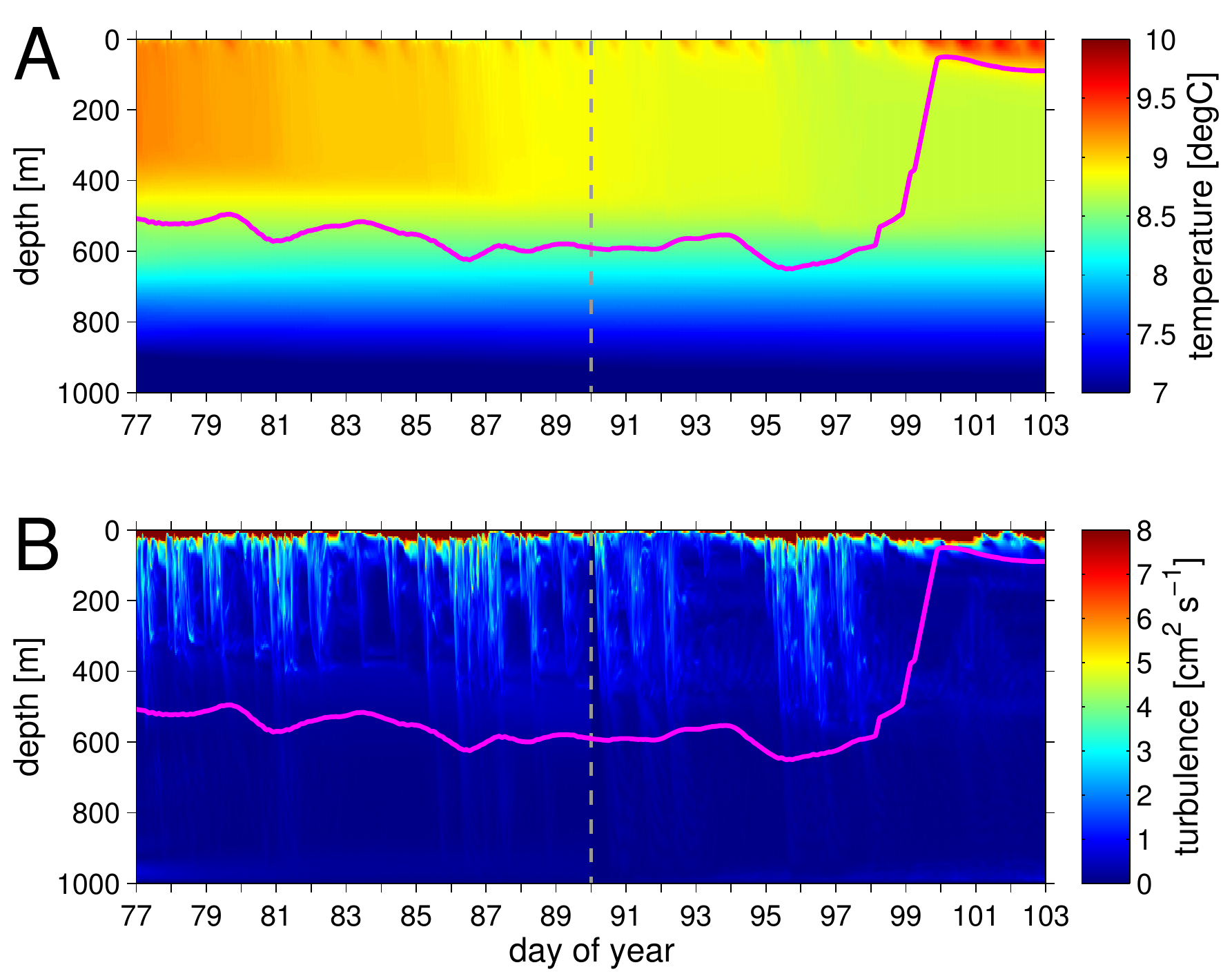}
\caption{Merged time series of simulated (A) temperature and (B) turbulence for the period from March 17th to April 12th. The upper end of the colour scale is cut at 8\,cm$^2$\,s$^{-1}$. Solid lines mark the daily running average of the MLD determined according to equation \eqref{equ:CML_crit} from the simulated temperature. Vertical dashed lines mark the date of assembly, March 30th.}
\label{fig:MLD&CML}
\end{figure}

\indent In order to validate our assumption that the MLD (after equation \eqref{equ:CML_crit}) is generally a good indicator for the convective mixing depth $H_{cml}$, we applied the convection model described in section \ref{subsec:ConvMod} to two different periods: March 17th to March 31st and March 30th to April 13th. \figref{fig:MLD&CML} shows the merged time series of simulated temperature (A) and turbulence as vertical mixing coefficient (B) from March 17th to April 12th. The colour scale for the turbulence is cut at its upper end at 8\,cm$^2$\,s$^{-1}$ to better resolve the values in greater depth. The vertical dashed lines (March 30th) mark the date of assembly. The daily running average of the MLD determined from the simulated temperature using the MLD criterion \eqref{equ:CML_crit} is marked by solid lines.\\
\indent The temperature shows a steady decrease within the upper 400\,m from the beginning of the simulation until April 2nd (day 93). This is in good agreement with the convective turnover times of 5 hours to 8 hours indicated by increased turbulence of about 2\,cm$^2$\,s\,$^{-1}$ down to depths of 300\,m to 400\,m during this period. The MLD in principle follows the temporal development of the turbulence, but the MLD is estimated about 150\,m deeper than the maximum depth of convective mixing during the whole simulation period. While the applied MLD criterion overestimates the $H_{cml}$, the MLD determined from the convection model and from the 3D simulation are in good agreement (see \figref{fig:TempHov}). The simulated turbulence also shows that the MLD is recurrently fully convected during winter, thus, providing a good indicator for the vertical extent of the convection cell.\\
\indent In the period after April 2nd, convection is strongly reduced. These dynamics are also visible in the temperature increase near the surface. During the same time, the MLD shows only a slight decline demonstrating that the MLD and the $H_{cml}$ deviate stronger during early spring. This period of three days of no convective mixing is followed by another period of strong convection causing a deepening of the mixed layer by almost 100\,m. The temperature within the whole mixed layer also significantly decreases during this time. Convection eventually drops on April 6th (day 97) followed by the formation of a shallow surface mixed layer. The onset of thermal stratification is delayed compared to the shutdown of convection by about two days. This shows that especially in the time directly after the shutdown of convection the MLD is not a good indicator for the $H_{cml}$ for the relevant dynamics to describe phytoplankton growth conditions. This is also in good agreement with \citet{taylor2011} and \citet{ferrari2014} who stated that during periods of strong atmospheric forcing the MLD is a good proxy for $H_{cml}$, whereas the MLD strongly deviates from $H_{cml}$ when the atmospheric forcing weakens in spring, thus, becoming a less strong driver for turbulent mixing. However, as our study focusses on primary production during winter when the MLD is frequently overturned by convection, we do not further discuss this problem here.\\
\indent The ratio $t_{exp}/t_{orb}$ is about 0.38 throughout the winter period (January - March) in the 3D phytoconvection run due to the spatial aspect ratio of 2.5:1 (horizontal to vertical) of the convection cell and an euphotic layer depth of 40\,m to 50\,m. To evaluate this value we calculated the median values of $t_{exp}$ and $t_{orb}$ for the 200 Lagrangian tracers added to the first simulation. The median was used rather than the average value to retrieve values less influenced by strong outliers occurring in the tracer results. We derived the values by considering tracers with an actual light value of above or equal 1\% of the surface light being within the euphotic zone, thus, counting for $t_{exp}$. Tracers with light values below this threshold were taken into account for the dark time $t_{dark}$. $t_{orb}$ was calculated as the sum of the median values of $t_{exp}$ and $t_{dark}$. The resulting median values for $t_{exp}$ and $t_{orb}$ are about 0.08 hours and about 0.42 hours, respectively, resulting in a ratio $t_{exp}/t_{orb}$ of about 0.2. This shows that the ratio $t_{exp}/t_{orb}$ during the 3D simulation is most likely overestimated causing an overestimation of the simulated phytoplankton. The short median duration of $t_{exp}$ and $t_{orb}$ suggests that particles in a convectively mixed layer are not always transported down to the bottom of the mixed layer, but spend a lot of time in the upper layers and frequently re-enter the euphotic zone. However, laboratory experiments showed that phytoplankton dealing with short light-dark cycles is less productive than that dealing with longer cycle due to photoacclimation (B. Walter, unpublished data). This is not taken into account in our parameterisation as the different light-dark cycles are not represented and would probably lower the simulated phytoplankton concentrations. However, all tracer trajectories (not presented) show that particles also spend longer continuous periods in different depths since convection is intermittent. Thus, it is likely that a certain proportion of phytoplankton stays sufficiently long in the euphotic zone and net primary production can take place.\\
\indent There are some differences between 2D and 3D modelling of turbulent convection, a point worth noting. For instance, \citet{moeng2004} showed that in the case of a free convective boundary layer as assumed by the model, the turbulent kinetic energy (TKE), surface friction velocity and velocity variances are sensitive to the subgrid-scale eddy viscosity and thermal diffusivity in 2D models. Especially the vertical velocity variance is significantly higher in 2D models than in 3D which affects the vertical mass fluxes \citep{moeng2004}. Hence, a 3D simulation using the same boundary conditions would lead to different results regarding the TKE, and thus, to a different $H_{cml}$ and MLD, respectively. \citet{foxkemper2008} showed that at the mixed layer base 2D models tend to simulate a lower vertical diffusivity compared to 3D models which may affect the temporal development of $H_{cml}$. Even though such model would most likely yield a quantitatively different result, this does not affect the validity of our assumption that the MLD equals the $H_{cml}$ during winter as we do not expect another qualitative result from a 3D model. With respect to the vertical displacement of particles by convection, the differences in the vertical velocities would affect the simulated duration of a full convective orbit, and thus, the ratio $t_{exp}/t_{orb}$.
\section{Conclusion and perspectives} \label{sec:Conclusions&Perspectives}
\indent The presented parameterisation of phytoconvection (equation \eqref{equ:PB_pc}) combined with the developed weighting function \eqref{equ:f_p} constitutes a possible approach to reproduce observed winter phytoplankton concentrations \citep[e.g.][]{backhaus2003} in a large-scale physical-biogeochemical model. In these models the vertical exchange is often not sufficient to account for convective mixing, and thus, to adequately simulate phytoplankton concentrations within the deep winter mixed layer. This parameterisation incorporates the hypothesis of \citet{backhaus1999} by allowing primary production throughout the whole mixed layer, and thus, increasing winter phytoplankton productivity. The resulting parameterisation (equation \eqref{equ:PB_total}) is able to simulate winter phytoplankton concentrations in convective regions which are in good agreement to the available observations whereas the conventional parameterisation of primary production fails to reproduce the observations. The more realistic winter phytoplankton concentrations and the accordingly higher carbon export production in the phytoconvection run suggest that convection non-resolving, biogeochemical models underestimate the carbon export as they also underestimate the phytoplankton concentrations in these depths, and hence, the sinking of particulate organic carbon.\\
\indent Beyond this, for the first time the parameterisation proposed by \citet{backhaus2003} was tested and implemented into a large-scale 3D biogeochemical ocean model. Thus, it was possible to investigate the effect of this parameterisation on the simulated ecosystem, for example, nutrients and zooplankton. Even though only primary production is directly influenced by the parameterisation, nutrients and zooplankton are also affected. However, the increase in zooplankton grazing and the reduction of nutrients in the pre-bloom phase only have a minor effect on the phytoplankton bloom development. Thus, these results question the relative importance of changes in the density-dependent grazing pressure during winter as formulated in the `disturbance-recoupling hypothesis' \citep{behrenfeld2013, behrenfeld2014}.\\
\indent In addition, the regional model ECOHAM4 served as a testing environment for the application of the parameterisation of phytoconvection in a global ocean model. The promising results with respect to phytoplankton concentrations suggest this step to investigate the effect on the phytoplankton and carbon export on a global scale.\\
\indent The parameterisation presented relies on a number of assumptions regarding the characteristics of convection and the functioning of primary production. First, it is assumed that the MLD is frequently mixed by convection during winter, thus, providing a valid measure for the vertical extent of convection. This was confirmed by the results of the convection model, although the applied MLD criterion overestimates the actual convective mixing depth $H_{cml}$. Second, the convection cell is assumed to have a rectangular shape with a horizontal to vertical aspect ratio of 2.5:1 \citep{backhaus2003} and particles moving with constant velocity along linear pathways. The median timescales of $t_{exp}$ and $t_{orb}$ resulting from the tracer experiment showed that this simplification leads to an overestimation of the ratio $t_{exp}/t_{orb}$. The short timescales also suggest that the actual motion within the convection cell is rather turbulent than linear which also effects the particle velocity. Horizontal current velocities in the ocean are usually larger than vertical velocities. Hence, assuming increased horizontal velocities would lead to a reduction of $t_{exp}$ and consequently lower primary production. The short light-dark cycles show that the simulated phytoplankton is biased by neglecting the effect of photoacclimation resulting in a further overestimation. Consequently, taking into account these aspects would most  likely result in an overall lower primary production during winter.\\
\indent Since primary production depends on absolute values of PAR, primary production taking place in depths deeper than the 1\%-threshold depth could partly balance this decrease in primary production as it would increase $t_{exp}$. Though it is not likely to have a strong effect during mid-winter when radiation is generally low, it may have a beneficial effect during March when surface radiation increases, and thus, primary production may occur in greater depth. Variation in respiration is another aspect not accounted for in the applied model. It is known that phytoplankton respiration rates decrease under low light conditions \citep[e.g.][]{falkowski1985}. Thus, including this effect is expected to increase the simulated NPP and, consequently, phytoplankton concentrations. However, we did not address these aspects and further studies are needed to analyse their influence on our parameterisation of phytoconvection.\\
\indent Besides this the analysis of the new parameterisation revealed some other points requiring improvement. Considering that the MLD is the main controlling factor in both the paramerisation of phytoconvection and in the transition between the winter and summer regimes, the use of a higher vertical resolution between 200\,m to 600\,m depth would improve its accuracy in spatially explicit, especially during the decline of the MLD. Such improvement would also be associated with increased computation times. There is also the need of improving the MLD criterion \eqref{equ:CML_crit} itself to account for haline stratification. This could be achieved by applying a density-based criterion following the algorithm developed by \citet{kara2000}. The comparison of the MLD criterion applied here with those based on a temperature difference of 0.8\,K proposed by \citet{kara2000} showed that these criteria yield even deeper MLDs which are less consistent with the simulated temperature distribution (see \figref{fig:TempHov}). Hence, further testing is needed to obtain the best-fitting criterion and to include the effect of salinity.\\
\indent The introduction of the weighting function \eqref{equ:f_p} allowed the smooth transition between winter and summer regimes with deep and shallow mixed layers, respectively. In addition, the weighting function produced improved results in the deep open ocean and shallow shelf regions. However, with respect to the transition from a deep to a shallow mixed layer, further improvement of the weighting between the conventional parameterisation and the parameterisation of phytoconvection is required.\\
\indent The simulations with the convection model showed that during phases of reduced convective mixing the MLD shows only a slight response, even though these phases last for a couple of days. The simulations also showed that there is a temporal mismatch between the final shutdown of convection and the decline of the MLD. Thus, a possible improvement to the weighting function \eqref{equ:f_p} as well as to the ratio $t_{exp}/t_{orb}$ (see equations \eqref{equ:Torb} and \eqref{equ:Texp}) would be the coupling to the $H_{cml}$ rather than to the MLD.\\
\indent The vertical exchange coefficients used in some hydrostatic models (as for example the HAMSOM model) in combination with the low vertical resolution are not always sufficient to account for convective mixing of particles. On the other hand, convection-resolving models are not feasible to large-scale models due to immense computation times, which represents the major challenge for the further improvement of the here presented parameterisation of phytoconvection. For this purpose, a possible solution could be the coupling of the parameterisation of phytoconvection to the air-sea heat flux, a proxy for convective mixing. \citet{taylor2011} used the net surface heat flux to show that the shutdown of convection can be a better indicator for the onset of the bloom than the MLD. However, other mechanisms, e.g. eddy-driven stratification \citep{mahadevan2012} can lead to stratification without a change in the net surface heat flux, and would, thus, still not be captured by this approach.\\
\indent Besides these possible and necessary improvements and the further analysis of the underlying assumptions, another important step is the application of the presented parameterisation in a model region where sufficient observation data are available to obtain a significant image of the quality of the parameterisation. Such can be achieved with the help of other large-scale, physical-biogeochemical models, used to calculate growth dependency on the local PAR in depth $z$. The parameterisation of phytoconvection presented here is based on the originally implemented parameterisation of light-dependent primary production, and thus, is expected to be applicable by the vast majority of this type of models.
\section{Acknowledgements}
\indent We thank Detlef Quadfasel for kindly providing us the ARGO float pictures and the British Oceanographic Data Centre (BODC) for providing the observation profile data. Furthermore, we thank Wilfried K{\"u}hn for proofreading the article and giving helpful comments. We also thank the reviewers for the constructive criticism. Fabian Gro{\ss}e thanks Bettina Walter for fruitful discussions. Christian Lindemann was partly financially supported by the FP7 programme EURO-BASIN. This study uses data from the Bangor University School of Ocean Sciences, provided by the British Oceanographic Data Centre and funded by the Land Ocean Interaction Study (LOIS).
%
%
\section{References}
\begingroup
\renewcommand{\section}[2]{}
\bibliographystyle{elsarticle-harv}

\endgroup
\end{document}